\documentclass[journal]{IEEEtran}

\usepackage[utf8]{inputenc}
\usepackage{amssymb}
\usepackage{latexsym}
\usepackage{siunitx}
\usepackage{xcolor}
\usepackage{colortbl}
\usepackage{amsmath}
\usepackage{nicefrac}
\usepackage{hyperref}
\usepackage{array}
\usepackage{caption}
\usepackage{footnote}   
\usepackage{sidecap}   
\newcolumntype{C}[1]{>{\centering\arraybackslash}p{#1}}
\newcolumntype{R}[1]{>{\raggedright\arraybackslash}p{#1}}

\definecolor{LightCyan}{rgb}{0.88,1,1}
\definecolor{LightGreen}{rgb}{0.67,1,0.9}
\definecolor{Gray}{gray}{0.9}

\usepackage{arydshln}
\usepackage{array,makecell}

\newcommand{\E}{\bf E}

\newcommand{\BB}{\textbf{B}}

\usepackage{cite}
\ifCLASSOPTIONcompsoc
  \usepackage[caption=false,font=normalsize,labelfont=sf,textfont=sf]{subfig}
\else
  \usepackage[caption=false,font=footnotesize]{subfig}
\fi

\usepackage{stfloats}

\ifCLASSINFOpdf
  \usepackage[pdftex]{graphicx}
  \DeclareGraphicsExtensions{.pdf,.jpeg,.png}
\else
\fi
\begin{document}
%
\title{Automatic segmentation with detection of local segmentation failures in cardiac MRI}
%
%

\author{Jörg~Sander,
        Bob D.~de Vos
        and~Ivana~Išgum
\thanks{J. Sander, B.D. de Vos and I. Išgum are with the Department of Biomedical Engineering and Physics, Amsterdam UMC, University of Amsterdam, Amsterdam, The Netherlands e-mail: \mbox{j.sander1@amsterdamumc.nl.}}
\thanks{J. Sander, B.D. de Vos and I. Išgum are with Amsterdam Cardiovascular Sciences, Amsterdam University Medical Center, University of Amsterdam, Amsterdam, The Netherlands}
\thanks{J. Sander and I. Išgum are also with Image Sciences Institute, University Medical \mbox{Center} Utrecht, Utrecht, The Netherlands}
\thanks{I. Išgum is with Department of Radiology and Nuclear Medicine, Amsterdam University Medical Center, University of Amsterdam, Amsterdam, The Netherlands}
}

\maketitle

\begin{abstract}
Segmentation of cardiac anatomical structures in cardiac magnetic resonance images (CMRI) is a prerequisite for automatic diagnosis and prognosis of cardiovascular diseases. To increase robustness and performance of segmentation methods this study combines automatic segmentation and assessment of segmentation uncertainty in CMRI to detect image regions containing local segmentation failures. Three state-of-the-art convolutional neural networks (CNN) were trained to automatically segment cardiac anatomical structures and obtain two measures of predictive uncertainty: entropy and a measure derived by MC-dropout. Thereafter, using the uncertainties another CNN was trained to detect local segmentation failures that potentially need correction by an expert. Finally, manual correction of the detected regions was simulated. Using publicly available CMR scans from the MICCAI 2017 ACDC challenge, the impact of CNN architecture and loss function for segmentation, and the uncertainty measure was investigated. Performance was evaluated using the Dice coefficient and 3D Hausdorff distance between manual and automatic segmentation. The experiments reveal that combining automatic segmentation with simulated manual correction of detected segmentation failures leads to statistically significant performance increase.
\end{abstract}


%
\IEEEpeerreviewmaketitle

\section{Introduction}


\IEEEPARstart{T}o perform diagnosis and prognosis of cardiovascular disease (CVD) medical experts depend on the reliable quantification of cardiac function \cite{white1987left}. Cardiac magnetic resonance imaging (CMRI) is currently considered the reference standard for quantification of ventricular volumes, mass and function \cite{grothues2002comparison}. Short-axis CMR imaging, covering the entire left and right ventricle (LV resp. RV) is routinely used to determine quantitative parameters of both ventricle's function. This requires manual or semi-automatic segmentation of corresponding cardiac tissue structures for end-diastole (ED) and end-systole (ES). 

Existing semi-automated or automated segmentation methods for CMRIs regularly require (substantial) manual intervention caused by lack of robustness. Manual or semi-automatic segmentation across a complete cardiac cycle, comprising \num{20} to \num{40} phases per patient, enables computation of parameters quantifying cardiac motion with potential diagnostic implications but due to the required workload, this is practically infeasible. Consequently, segmentation is often performed at end-diastole and end-systole precluding comprehensive analysis over complete cardiac cycle. 

Recently\cite{litjens2017survey, leiner2019machine}, deep learning segmentation methods have shown to outperform traditional approaches such as those exploiting level set, graph-cuts, deformable models, cardiac atlases and statistical models \cite{petitjean2011review, peng2016review}. However, recent comparison of a number of automatic methods showed that even the best performing methods generated anatomically implausible segmentations in more than 80\% of the CMRIs \cite{bernard2018deep}. Such errors do not occur when experts  perform segmentation. To achieve acceptance in clinical practice these shortcomings of the automatic approaches need to be alleviated by further development. This can be achieved by generating more accurate segmentation result or by development of approaches that automatically detect segmentation failures. 

In manual and automatic segmentation of short-axis CMRI, largest segmentation inaccuracies are typically located in the most basal and apical slices due to low tissue contrast ratios \cite{suinesiaputra2015quantification}. To increase segmentation performance, several methods have been proposed \cite{tan2018fully, zheng20183, savioli2018automated, bai2018automated}. Tan et al. \cite{tan2018fully} used a convolutional neural network (CNN) to regress anatomical landmarks from long-axis views (orthogonal to short-axis). They exploited the landmarks to determine most basal and apical slices in short-axis views and thereby constraining the automatic segmentation of CMRIs. This resulted in increased robustness and performance. Other approaches leverage spatial \cite{zheng20183} or temporal \cite{savioli2018automated, bai2018automated} information to increase segmentation consistency and performance in particular in the difficult basal and apical slices.

An alternative approach to preventing implausible segmentation results is by incorporating knowledge about the highly constrained shape of the heart. Oktay et al. \cite{oktay2017anatomically} developed an anatomically constrained neural network (NN) that infers shape constraints using an auto-encoder during segmentation training. Duan et al. \cite{duan2019automatic} developed a deep learning segmentation approach for CMRIs that used atlas propagation to explicitly impose a shape refinement. This was especially beneficial in the presence of image acquisition artifacts. Recently, Painchaud et al. \cite{painchaud2019cardiac} developed a post-processing approach to detect and transform anatomically implausible cardiac segmentations into valid ones by defining cardiac anatomical metrics. Applying their approach to various state-of-the-art segmentation methods the authors showed that the proposed method provides strong anatomical guarantees without hampering segmentation accuracy. 

\begin{figure*}[t]
	\center
	\includegraphics[width=6in, height=3.5in]{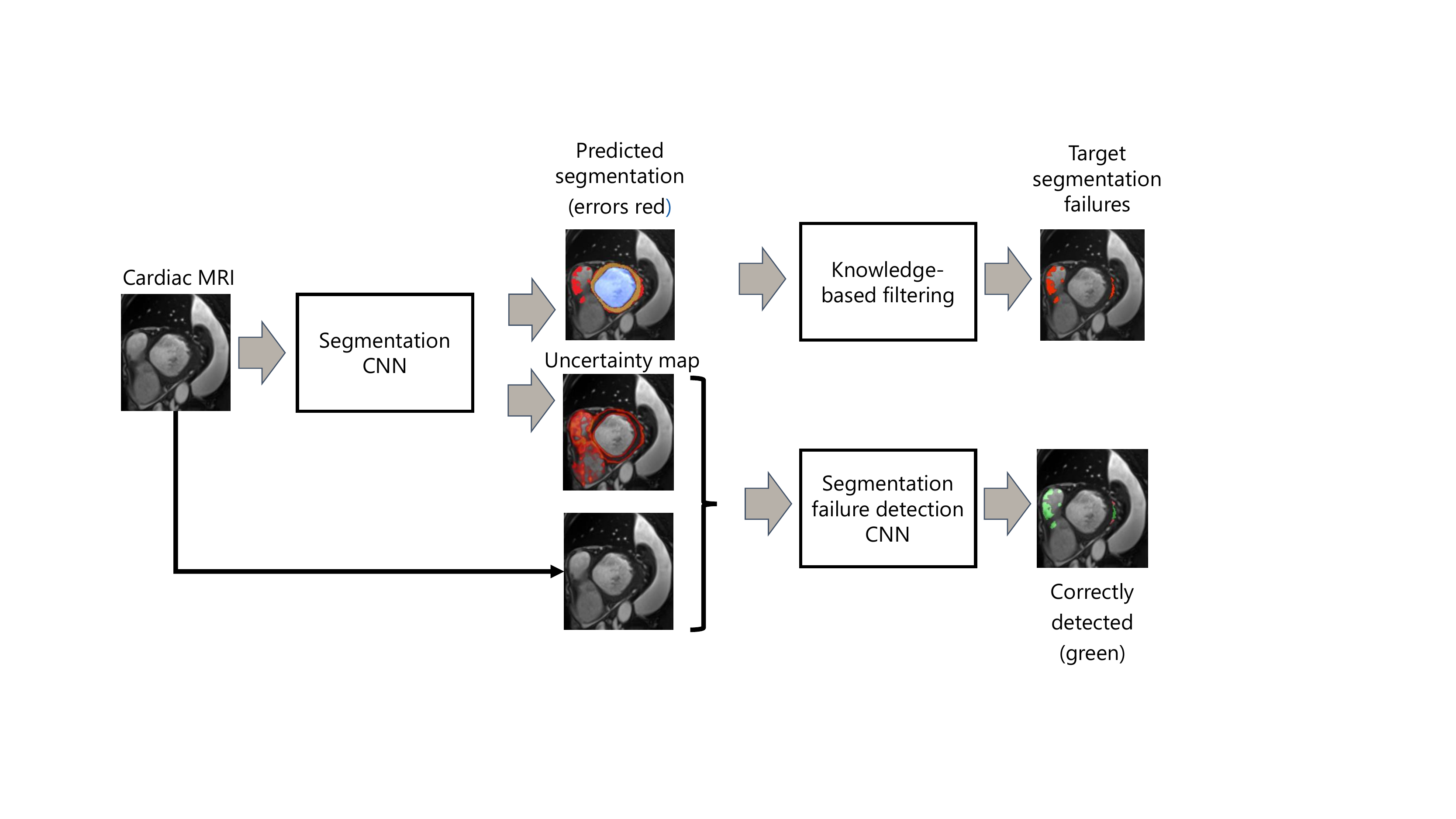}%
	
	\caption{Overview of proposed two step approach. Step 1 (left): Automatic CNN segmentation of CMR images combined with assessment of segmentation uncertainties. Step 2 (right): Differentiate tolerated errors from segmentation failures (to be detected) using distance transform maps based on reference segmentations. Detection of image regions containing segmentation failures using CNN which takes CMR images and segmentation uncertainties as input. Manual corrected segmentation failures (green) based on detected image regions.}
	\label{fig_overview_method}
\end{figure*}

A different research trend focuses on detecting segmentation failures, i.e. on automated quality control for image segmentation. These methods can be divided in those that predict segmentation quality using image at hand or corresponding automatic segmentation result, and those that assess and exploit predictive uncertainties to detect segmentation failure. 

Recently, two methods were proposed to detect segmentation failures in large-scale cardiac MR imaging studies to remove these from subsequent analysis\cite{alba2018automatic, robinson2019automated}. Robinson et al. \cite{robinson2019automated} using the approach of Reverse Classification Accuracy (RCA) \cite{valindria2017reverse} predicted CMRI segmentation metrics to detect failed segmentations. They achieved good agreement between predicted metrics and  visual quality control scores. Alba et al. \cite{alba2018automatic} used statistical, pattern and fractal descriptors in a random forest classifier to directly detect segmentation contour failures without intermediate regression of segmentation accuracy metrics. 

Methods for automatic quality control were also developed for other applications in medical image analysis. Frounchi et al. \cite{frounchi2011automating} extracted features from the segmentation results of the left ventricle in CT scans. Using the obtained features the authors trained a classifier that is able to discriminate between consistent and inconsistent segmentations. To distinguish between acceptable and non-acceptable segmentations Kohlberger el al. \cite{kohlberger2012evaluating} proposed to directly predict multi-organ segmentation accuracy in CT scans using a set of features extracted from the image and corresponding segmentation. 

A number of methods aggregate voxel-wise uncertainties into an overall score to identify insufficiently accurate segmentations. For example, Nair et al. \cite{nair2018exploring} computed an overall score for target segmentation structure from voxel-wise predictive uncertainties. The method was tested for detection of Multiple Sclerosis in brain MRI. The authors showed that rejecting segmentations with high uncertainty scores led to increased detection accuracy indicating that correct segmentations contain lower uncertainties than incorrect ones. Similarly, to assess segmentation quality of brain MRIs Jungo et al. \cite{jungo2018uncertainty} aggregated voxel-wise uncertainties into a score per target structure 
and showed that the computed uncertainty score enabled identification of erroneous segmentations.

Unlike approaches evaluating segmentation directly, several methods use predictive uncertainties to predict segmentation metrics and thereby evaluate segmentation performance \cite{roy2019bayesian, devries2018leveraging}. For example, Roy et al. \cite{roy2019bayesian} aggregated voxel-wise uncertainties into four scores per segmented structure in brain MRI. The authors showed that computed scores can be used to predict the Intersection over Union and hence, to determine segmentation accuracy. 
Similar idea was presented by DeVries et al. \cite{devries2018leveraging} that predicted segmentation accuracy per patient using an auxiliary neural network that leverages the dermoscopic image, automatic segmentation result and obtained uncertainties. The researchers showed that a predicted segmentation accuracy is useful for quality control.

We build on our preliminary work where automatic segmentation of CMR images using a dilated CNN was combined with assessment of two measures of segmentation uncertainties \cite{sander2019towards}. For the first measure the multi-class entropy per voxel (entropy maps) was computed using the output distribution. For the second measure Bayesian uncertainty maps were acquired using Monte Carlo dropout (MC-dropout) \cite{gal2016dropout}. In \cite{sander2019towards} we showed that the obtained uncertainties almost entirely cover the regions of incorrect segmentation i.e. that uncertainties are calibrated. In the current work we extend our preliminary research in two ways. First, we assess impact of CNN architecture on the segmentation performance and calibration of uncertainty maps by evaluating three existing state-of-the-art CNNs. Second, we employ an auxiliary CNN (detection network) that processes a cardiac MRI and corresponding spatial uncertainty map (Entropy or Bayesian) to automatically detect segmentation failures. We differentiate errors that may be within the range of inter-observer variability and hence do not necessarily require correction (tolerated errors) from the errors that an expert would not make and hence require correction (segmentation failures). Given that overlap measures do not capture fine details of the segmentation results and preclude us to differentiate two types of segmentation errors, in this work, we define segmentation failure using a metric of boundary distance. In \cite{sander2019towards} we found that degree of calibration of uncertainty maps is dependent on the loss function used to train the CNN. Nevertheless, in the current work we show that uncalibrated uncertainty maps are useful to detect local segmentation failures. 	
In contrast to previous methods that detect segmentation failure per-patient or per-structure\cite{roy2019bayesian, devries2018leveraging}, we propose to detect segmentation failures per image region. We expect that inspection and correction of segmentation failures using image regions rather than individual voxels or images would simplify correction process. To show the potential of our approach and demonstrate that combining automatic segmentation with manual correction of the detected segmentation failures per region results in higher segmentation performance we performed two additional experiments. In the first experiment, correction of detected segmentation failures was simulated in the complete data set. In the second experiment, correction was performed by an expert in a subset of images. Using publicly available set of CMR scans from MICCAI 2017 ACDC challenge \cite{bernard2018deep}, the performance was evaluated before and after simulating the correction of detected segmentation failures as well as after manual expert correction.

\section{Data}

In this study data from the MICCAI \num{2017} Automated Cardiac Diagnosis Challenge (ACDC) \cite{bernard2018deep} was used. The dataset consists of cardiac cine MR images (CMRIs) from 100 patients uniformly distributed over normal cardiac function and four disease groups: dilated cardiomyopathy, hypertrophic cardiomyopathy, heart failure with infarction, and right ventricular abnormality. Detailed acquisition protocol is described by Bernard et al.~\cite{bernard2018deep}. Briefly, short-axis CMRIs were acquired with two MRI scanners of different magnetic strengths (\num{1.5} and \num{3.0} T). Images were made during breath hold using a conventional steady-state free precession (SSFP) sequence. CMRIs have an in-plane resolution ranging from \num{1.37} to \SI{1.68}{\milli\meter} (average reconstruction matrix \num{243} $\times$ \num{217} voxels) with slice spacing varying from \num{5} to \SI{10}{\milli\meter}. Per patient 28 to 40 volumes  are provided covering partially or completely one cardiac cycle. Each volume consists of on average ten slices covering the heart. Expert manual reference segmentations are provided for the LV cavity, RV endocardium and LV myocardium (LVM) for all CMRI slices at ED and ES time frames. To correct for intensity differences among scans, voxel intensities of each volume were scaled to the [\num{0.0}, \num{1.0}] range using the minimum and maximum of the volume. Furthermore, to correct for differences in-plane voxel sizes, image slices were resampled to \num{1.4}$\times\SI{1.4}{\milli\meter}^2$. 

\begin{figure*}[!t]
	\captionsetup[subfigure]{justification=centering}
	\centering
	\subfloat[]{\includegraphics[width=5in]{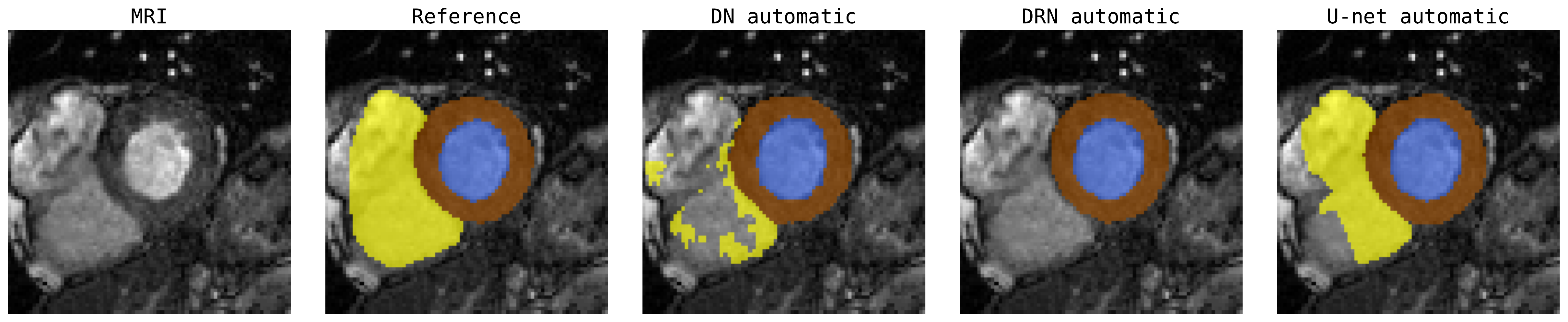}%
		\label{fig_seg_qual_example1}}
	
	\subfloat[]{\includegraphics[width=5in]{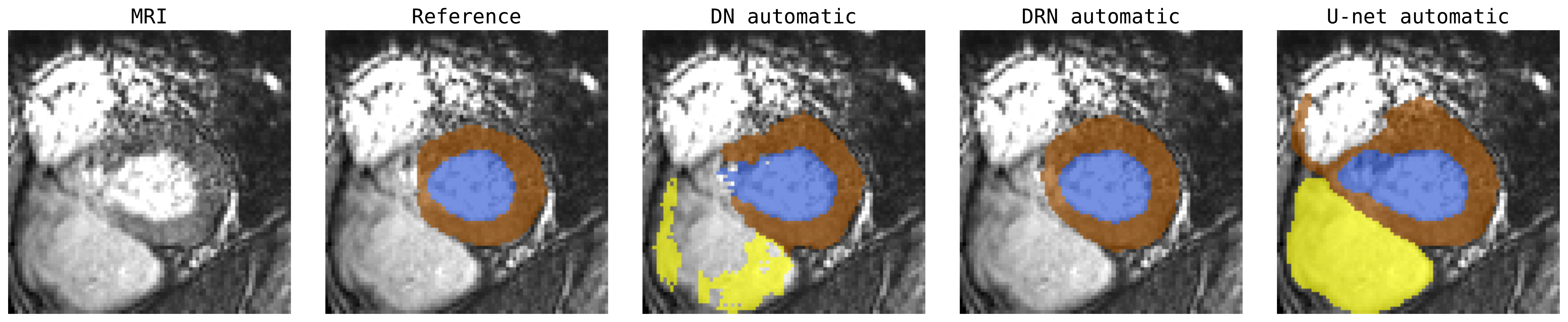}%
		\label{fig_seg_qual_example2}}
	
	\caption{Examples of automatic segmentations generated by different segmentation models for two cardiac MRI scans (rows) at ES at the base of the heart.}
	\label{fig_seg_qualitative_results}
\end{figure*}

\section{Methods}

To investigate uncertainty of the segmentation, anatomical structures in CMR images are segmented using a CNN. To investigate whether the approach generalizes to different segmentation networks, three state-of-the-art CNNs were evaluated. For each segmentation model two measures of predictive uncertainty were obtained per voxel. Thereafter, to detect and correct local segmentation failures an auxiliary CNN (detection network) that analyzes a cardiac MRI was used. Finally, this leads to the uncertainty map allowing detection of image regions that contain segmentation failures. Figure~\ref{fig_overview_method} visualizes this approach.

\subsection{Automatic segmentation of cardiac MRI}

To perform segmentation of LV, RV, and LVM in cardiac MR images i.e. \num{2}D CMR scans, three state-of-the-art CNNs are trained. Each of the three networks takes a CMR image as input and has four output channels providing probabilities for the three cardiac structures (LV, RV, LVM) and background. Softmax probabilities are calculated over the four tissue classes. Patient volumes at ED and ES are processed separately. During inference the \num{2}D automatic segmentation masks are stacked into a \num{3}D volume per patient and cardiac phase. After segmentation, the largest \num{3}D connected component for each class is retained and volumes are resampled to their original voxel resolution. Segmentation networks differ substantially regarding architecture, number of parameters and receptive field size. To assess predictive uncertainties from the segmentation models \textit{Monte Carlo dropout} (MC-dropout) introduced by Gal \& Ghahramani \cite{gal2016dropout} is implemented in every network. The following three segmentation networks were evaluated: Bayesian Dilated CNN, Bayesian Dilated Residual Network, Bayesian U-net.

\vspace{1ex}
\noindent \textbf{Bayesian Dilated CNN (DN)}: The Bayesian DN architecture comprises a sequence of ten convolutional layers. Layers \num{1} to \num{8} serve as feature extraction layers with small convolution kernels of size \num{3}$\times$\num{3} voxels. No padding is applied after convolutions. The number of kernels increases from \num{32} in the first eight layers, to \num{128} in the final two fully connected classification layers, implemented as \num{1}$\times$\num{1} convolutions. The dilation level is successively increased between layers \num{2} and \num{7} from \num{2} to \num{32} which results in a receptive field for each voxel of \num{131}$\times$\num{131} voxels, or \num{18.3}$\times$ $\SI{18.3}{\centi\meter}^2$. All trainable layers except the final layer use rectified linear activation functions (ReLU). To enhance generalization performance, the model uses batch normalization in layers \num{2} to \num{9}. In order to convert the original DN~\cite{wolterink2017automatic} into a Bayesian DN dropout is added as the last operation in all but the final layer and \num{10} percent of a layer's hidden units are randomly switched off. 

\vspace{1ex}
\noindent \textbf{Bayesian Dilated Residual Network (DRN)}: The Bayesian DRN is based on the original DRN from Yu et al. \cite{yu2017dilated} for image segmentation. More specifically, the DRN-D-22\cite{yu2017dilated} is used which consists of a feature extraction module with output stride eight followed by a classifier implemented as fully convolutional layer with \num{1}$\times$\num{1} convolutions. Output of the classifier is upsampled to full resolution using bilinear interpolation. The convolutional feature extraction module comprises eight levels where the number of kernels increases from \num{16} in the first level, to \num{512} in the two final levels. The first convolutional layer in level \num{1} uses \num{16} kernels of size \num{7}$\times$\num{7} voxels and zero-padding of size \num{3}. The remaining trainable layers use small \num{3}$\times$\num{3} voxel kernels and zero-padding of size \num{1}. Level \num{2} to \num{4} use a strided convolution of size \num{2}. To further increase the receptive field convolutional layers in level \num{5}, \num{6} and \num{7} use a dilation factor of \num{2}, \num{4} and \num{2}, respectively. Furthermore, levels \num{3} to \num{6} consist of two residual blocks. All convolutional layers of the feature extraction module are followed by batch normalization, ReLU function and dropout. Adding dropout and switching off \num{10} percent of a layer's hidden units converts the original DRN~\cite{yu2017dilated} into a Bayesian DRN.

\vspace{1ex}
\noindent \textbf{Bayesian U-net (U-net)}: The standard architecture of the U-net~\cite{ronneberger2015u} is used. The network is fully convolutional and consists of a contracting, bottleneck and expanding path. The contracting and expanding path each consist of four blocks i.e. resolution levels which are connected by skip connections. The first block of the contracting path contains two convolutional layers using a kernel size of \num{3}$\times$\num{3} voxels and zero-padding of size \num{1}. Downsampling of the input is accomplished by employing a max pooling operation in block \num{2} to \num{4} of the contracting path and the bottleneck using a convolutional kernel of size \num{2}$\times$\num{2} voxels and stride \num{2}. Upsampling is performed by a transposed convolutional layer in block \num{1} to \num{4} of the expanding path using the same kernel size and stride as the max pooling layers. Each downsampling and upsampling layer is followed by two convolutional layers using \num{3}$\times$\num{3} voxel kernels with zero-padding size \num{1}. The final convolutional layer of the network acts as a classifier and uses \num{1}$\times$\num{1} convolutions to reduce the number of output channels to the number of segmentation classes. The number of kernels increases from \num{64} in the first block of the contracting path to \num{1024} in the bottleneck. In contrast, the number of kernels in the expanding path successively decreases from \num{1024} to \num{64}. In deviation to the standard U-net instance normalization is added to all convolutional layers in the contracting path and ReLU non-linearities are replaced by LeakyReLU functions because this was found to slightly improve segmentation performance. In addition, to convert the deterministic model into a Bayesian neural network dropout is added as the last operation in each block of the contracting and expanding path and \num{10} percent of a layer's hidden units are randomly switched off.

\subsection{Assessment of predictive uncertainties} \label{uncertainty_maps}
To detect failures in segmentation masks generated by CNNs in testing, spatial uncertainty maps of the obtained segmentations are generated. For each voxel in the image two measures of uncertainty are calculated. First, a computationally cheap and straightforward measure of uncertainty is the entropy of softmax probabilities over the four tissue classes which are generated by the segmentation networks. Using these, normalized entropy maps $\E \in [0, 1]^{H\times W}$ (e-map) are computed where $H$ and $W$ denote the height and width of the original CMRI, respectively.

Second, by applying MC-dropout in testing, softmax probabilities with a number of samples $T$ per voxel are obtained. As an overall measure of uncertainty the mean standard deviation of softmax probabilities per voxel over all tissue classes $C$\label{ref:maximum_variance} is computed

\begingroup
\small
\begin{align}
\BB (I)^{(x, y)} &= \frac{1}{C} \sum_{c=1}^{C} \sqrt{\frac{1}{T-1}  \sum_{t=1}^{T} \big(p_t(I)^{(x, y, c)} - \hat{\mu}^{(x, y, c)} \big)^2 } \; ,
\end{align}
\endgroup

where $\BB(I)^{(x, y)} \in [0, 1]$ denotes the normalized value of the Bayesian uncertainty map (b-map) at position $(x, y)$ in \num{2}D slice $I$, $C$ is equal to the number of classes, $T$ is the number of samples and $p_t(I)^{(x, y, c)}$ denotes the softmax probability at position $(x, y)$  in image $I$ for class $c$. The predictive mean per class $\hat{\mu}^{(x, y, c)}$ of the samples is computed as follows:

\begingroup
\small
\begin{align}
\hat{\mu}^{(x, y, c)} &= \frac{1}{T} \sum_{t=1}^{T} p_t(I)^{(x, y, c)} \; .
\end{align}
\endgroup

In addition, the predictive mean per class is used to determine the tissue class per voxel.

\subsection{Calibration of uncertainty maps} 
Ideally, incorrectly segmented voxels as defined by the reference labels should be covered by higher uncertainties than correctly segmented voxels. In such a case the spatial uncertainty maps are perfectly calibrated. \textit{Risk-coverage curves} introduced by Geifman et al.\cite{geifman2017selective} visualize whether incorrectly segmented voxels are covered by higher uncertainties than those that are correctly segmented. Risk-coverage curves convey the effect of avoiding segmentation of voxels above a specific uncertainty value on the reduction of segmentation errors (i.e. risk reduction) while at the same time quantifying the voxels that were omitted from the classification task (i.e. coverage). 

To generate risk-coverage curves first, each patient volume is cropped based on a minimal enclosing  parallelepiped bounding box that is placed around the reference segmentations to reduce the number of background voxels. Note that this is only performed to simplify the analysis of the risk-coverage curves. Second, voxels of the cropped patient volume are ranked based on their uncertainty value in descending order. Third, to obtain uncertainty threshold values per patient volume the ranked voxels are partitioned into \num{100} percentiles based on their uncertainty value. Finally, per patient volume each uncertainty threshold is evaluated by computing a coverage and a risk measure. Coverage is the percentage of voxels in a patient volume at ED or ES that is automatically segmented. Voxels in a patient volume above the threshold are discarded from automatic segmentation and would be referred to an expert. The number of incorrectly segmented voxels per patient volume is used as a measure of risk. Using bilinear interpolation risk measures are computed per patient volume between $[0, 100]$ percent.

\subsection{Detection of segmentation failures}

To detect segmentation failures uncertainty maps are used but direct application of uncertainties is infeasible because many correctly segmented voxels, such as those close to anatomical structure boundaries, have high uncertainty. Hence, an additional patch-based CNN (detection network) is used that takes a cardiac MR image together with the corresponding spatial uncertainty map as input. For each patch of \num{8}$\times$\num{8} voxels the network generates a probability indicating whether it contains segmentation failure. In the following, the terms patch and region are used interchangeably.

The detection network is a shallow Residual Network (S-ResNet) \cite{he2016deep} consisting of a feature extraction module with output stride eight followed by a classifier indicating the presence of segmentation failure. The first level of the feature extraction module consists of two convolutional layers. The first layer uses \num{16} kernels of \num{7}$\times$\num{7} voxels and zero-padding of size \num{3} and second layer \num{32} kernels of \num{3}$\times$\num{3} voxels and zero-padding of \num{1} voxel. Level \num{2} to \num{4} each consist of one residual block that contains two convolutional layers with \num{3}$\times$\num{3} voxels kernels with zero-padding of size \num{1}. The first convolutional layer of each residual block uses a strided convolution of \num{2} voxels to downsample the input. All convolutional layers of the feature extraction module are followed by batch normalization and ReLU function. The number of kernels in the feature extraction module increases from \num{16} in level \num{1} to \num{128} in level \num{4}. The network is a \num{2}D patch-level classifier and requires that the size of the two input slices is a multiple of the patch-size. \label{patch_size} The final classifier consists of three fully convolutional layers, implemented as \num{1}$\times$\num{1} convolutions, with \num{128} feature maps in the first two layers. The final layer has two channels followed by a softmax function which indicates whether the patch contains segmentation failure. Furthermore, to regularize the model dropout layers ($p=0.5$) were added between the residual blocks and the fully convolutional layers of the classifier.

\begin{table*}
	\caption{Segmentation performance of different combination of model architectures, loss functions and evaluation modes (without or with MC dropout enabled during testing) in terms of Dice coefficient (top) and Hausdorff distance (bottom) (mean $\pm$ standard deviation). Each combination comprises a block of two rows. A row in which column \textit{Uncertainty map for detection} indicates e- or b-map shows results for the combined segmentation and detection approach. Numbers accentuated in black/bold are ranked first in the segmentation only task whereas numbers accentuated in red/bold are ranked first in the combined segmentation \& detection task. The last row states the performance of the winning model in the ACDC challenge (on \num{100} patient images) \cite{isensee2017automatic}. Number with asterisk indicates statistical significant at \num{5}\% level w.r.t. the segmentation-only approach. Best viewed in color.}
	\label{table_overall_segmentation_performance}
	\centering
	\tiny
	\subfloat[Dice coefficient]{
		\begin{tabular}{| C{1.6cm} | C{0.8cm} | R{1.7cm} R{1.7cm} R{1.7cm} | R{1.7cm} R{1.7cm} R{1.7cm} | }
			\hline
			& \textbf{Uncertainty} & \multicolumn{3}{c|}{\textbf{End-diastole}} & \multicolumn{3}{c|}{\textbf{End-systole}} \\
			\textbf{Model} & \textbf{map for detection} & \multicolumn{1}{l}{\textbf{LV}} & \multicolumn{1}{l}{\textbf{RV}} & \multicolumn{1}{l|}{\textbf{LVM}} & \multicolumn{1}{l}{\textbf{LV}} & \multicolumn{1}{l}{\textbf{RV}} & \multicolumn{1}{l|}{\textbf{LVM}} \\ 
			\hline
			\rowcolor{LightGreen}
			DN-Brier &  & \phantom{x}0.962$\pm$0.02 & \phantom{x}0.928$\pm$0.04) & \phantom{x}0.875$\pm$0.03)  & \phantom{x}0.901$\pm$0.11 & \phantom{x}0.832$\pm$0.10) & \phantom{x}0.884$\pm$0.04 \\
			& \textbf{e-map} & *0.965$\pm$0.01 & *0.949$\pm$0.02 & *0.885$\pm$0.03  & *0.937$\pm$0.06 & *0.905$\pm$0.05 & *0.909$\pm$0.03 \\ 
			
			\hdashline[1pt/2pt]
			
			\rowcolor{LightCyan}
			DN-Brier+MC  & & \phantom{x}0.961$\pm$0.02 & \phantom{x}0.922$\pm$0.04 & \phantom{x}0.875$\pm$0.04 & \phantom{x}0.912$\pm$0.08  & \phantom{x}0.839$\pm$0.11 & \phantom{x}0.882$\pm$0.04  \\ 
			& \textbf{b-map} & *0.966$\pm$0.01 & *0.950$\pm$0.01 & *0.886$\pm$0.03  & \textbf{\textcolor{red}{*0.942}}$\pm$0.03 & *0.916$\pm$0.04 & *0.912$\pm$0.03 \\ 
			\hdashline[5pt/5pt]
			
			\rowcolor{LightGreen}
			DN-soft-Dice  & & \phantom{x}0.960$\pm$0.02 & \phantom{x}0.921$\pm$0.04 & \phantom{x}0.870$\pm$0.04 &  \phantom{x}0.909$\pm$0.08 & \phantom{x}0.812$\pm$0.12 & \phantom{x}0.879$\pm$0.04 \\
			& \textbf{e-map} & *0.965$\pm$0.01 & *0.945$\pm$0.02 & *0.879$\pm$0.04  & *0.938$\pm$0.03 & *0.891$\pm$0.06 & *0.905$\pm$0.03 \\ 
			\hdashline[1pt/2pt]
			
			\rowcolor{LightCyan}
			DN-soft-Dice+MC & & \phantom{x}0.958$\pm$0.02 & \phantom{x}0.913$\pm$0.05  & \phantom{x}0.868$\pm$0.04 & \phantom{x}0.907$\pm$0.07  & \phantom{x}0.818$\pm$0.12 & \phantom{x}0.875$\pm$0.04  \\
			& \textbf{b-map} & *0.964$\pm$0.01 & *0.944$\pm$0.02 & *0.877$\pm$0.04  & *0.939$\pm$0.03 & *0.900$\pm$0.05 & *0.904$\pm$0.03 \\  			
			\hdashline[5pt/5pt]
			
			\rowcolor{LightGreen}
			DRN-CE &  & \phantom{x}0.961$\pm$0.02 & \phantom{x}0.929$\pm$0.03 & \phantom{x}0.878$\pm$0.03  & \phantom{x}0.912$\pm$0.06 & \phantom{x}0.850$\pm$0.09 & \phantom{x}0.891$\pm$0.03 \\  
			& \textbf{e-map} & \phantom{x}0.964$\pm$0.01 & *0.943$\pm$0.02 & *0.886$\pm$0.03  & *0.937$\pm$0.03 & *0.899$\pm$0.04 & *0.908$\pm$0.03 \\
			\hdashline[1pt/2pt]

			\rowcolor{LightCyan}
			DRN-CE+MC &  & \phantom{x}0.961$\pm$0.02 & \phantom{x}0.926$\pm$0.03 & \phantom{x}0.877$\pm$0.03  & \phantom{x}0.913$\pm$0.06 & \phantom{x}0.847$\pm$0.10 & \phantom{x}0.890$\pm$0.03 \\  
			& \textbf{b-map} & *0.965$\pm$0.01 & *0.948$\pm$0.01 & *0.887$\pm$0.03  & *0.939$\pm$0.03 & *0.911$\pm$0.04 & *0.909$\pm$0.03 \\ 
			\hdashline[5pt/5pt]

			\rowcolor{LightGreen}
			DRN-soft-Dice &  & \phantom{x}0.964$\pm$0.01 & \phantom{x}\textbf{0.937}$\pm$0.02 & \phantom{x}0.888$\pm$0.03  & \phantom{x}\textbf{0.919}$\pm$0.06 & \phantom{x}0.856$\pm$0.09 & \phantom{x}\textbf{0.900}$\pm$0.03 \\ 
			& \textbf{e-map} & \phantom{x}0.967$\pm$0.01 & *0.945$\pm$0.02 & \phantom{x}0.893$\pm$0.03  & \phantom{x}0.934$\pm$0.04 & *0.892$\pm$0.06 & *0.911$\pm$0.03 \\ 
			\hdashline[1pt/2pt]

			\rowcolor{LightCyan}
			DRN-soft-Dice+MC &  & \phantom{x}0.963$\pm$0.02 & \phantom{x}0.935$\pm$0.03 & \phantom{x}0.886$\pm$0.03  & \phantom{x}0.921$\pm$0.06 & \phantom{x}\textbf{0.857}$\pm$0.09 & \phantom{x}0.899$\pm$0.03 \\ 
			& \textbf{b-map} & \phantom{x}0.967$\pm$0.01 & *0.947$\pm$0.02 & \phantom{x}0.893$\pm$0.03  & *0.938$\pm$0.03 & *0.907$\pm$0.04 & *0.912$\pm$0.03 \\
			\hdashline[5pt/5pt]
			
			\rowcolor{LightGreen}
			U-net-CE &  & \phantom{x}0.962$\pm$0.02 & \phantom{x}0.923$\pm$0.05 & \phantom{x}0.878$\pm$0.03  & \phantom{x}0.907$\pm$0.07 & \phantom{x}0.840$\pm$0.08 & \phantom{x}0.885$\pm$0.03 \\ 
			& \textbf{e-map} & \phantom{x}0.966$\pm$0.01 & *0.946$\pm$0.02 & *0.890$\pm$0.03  & *0.935$\pm$0.04 & *0.901$\pm$0.06 & *0.909$\pm$0.03 \\ 
			
			\hdashline[1pt/2pt]
			
			\rowcolor{LightCyan}
			U-net-CE+MC &  & \phantom{x}0.962$\pm$0.02 & \phantom{x}0.926$\pm$0.04 & \phantom{x}0.879$\pm$0.03  & \phantom{x}0.909$\pm$0.07 & \phantom{x}0.849$\pm$0.07 & \phantom{x}0.887$\pm$0.03 \\ 
			& \textbf{b-map} & \phantom{x}0.967$\pm$0.01 & \textbf{\textcolor{red}{*0.954}}$\pm$0.02 & *0.893$\pm$0.03  & *0.940$\pm$0.04 & \textbf{\textcolor{red}{*0.920}}$\pm$0.04 & \textbf{\textcolor{red}{*0.914}}$\pm$0.03 \\
			\hdashline[5pt/5pt]
			
			\rowcolor{LightGreen}
			U-net-soft-Dice &  & \phantom{x}\textbf{0.965}$\pm$0.02 & \phantom{x}0.928$\pm$0.04 & \phantom{x}0.888$\pm$0.03  & \phantom{x}0.914$\pm$0.08 & \phantom{x}0.844$\pm$0.09 & \phantom{x}0.896$\pm$0.03 \\ 
			& \textbf{e-map} & \phantom{x}0.968$\pm$0.01 & *0.943$\pm$0.03 & *0.898$\pm$0.03  & \phantom{x}0.930$\pm$0.05 & *0.886$\pm$0.07 & *0.911$\pm$0.03 \\ 
			\hdashline[1pt/2pt]
			
			\rowcolor{LightCyan}
			U-net-soft-Dice+MC &  & \phantom{x}\textbf{0.965}$\pm$0.02 & \phantom{x}0.929$\pm$0.04 & \phantom{x}\textbf{0.889}$\pm$0.03  & \phantom{x}0.911$\pm$0.10 & \phantom{x}0.845$\pm$0.09 & \phantom{x}0.897$\pm$0.03 \\ 
			& \textbf{b-map} & \phantom{x}\textbf{\textcolor{red}{0.968}}$\pm$0.01 & *0.948$\pm$0.03 & \textbf{\textcolor{red}{*0.900}}$\pm$0.03  & \phantom{x}0.928$\pm$0.09 & *0.895$\pm$0.06 & \textbf{\textcolor{red}{*0.914}}$\pm$0.03 \\
			
			\hdashline
			Isensee et al. & & \phantom{x}0.966& \phantom{x}0.941 & \phantom{x}0.899 & \phantom{x}0.924 & \phantom{x}0.875	& \phantom{x}0.908  \\
			
			\hline
			
		\end{tabular}
		\label{table_seg_perf_dsc}
	}  
	\vspace{13ex}
	\centering
	\tiny
	\subfloat[Hausdorff Distance]{
		\begin{tabular}{| C{1.6cm} | C{0.8cm} | R{1.7cm} R{1.7cm} R{1.7cm} | R{1.7cm} R{1.7cm} R{1.7cm} | }
			\hline
			& \textbf{Uncertainty}& \multicolumn{3}{c|}{\textbf{End-diastole}} & \multicolumn{3}{c|}{\textbf{End-systole}} \\
			\textbf{Model} & \textbf{map for detection} & \multicolumn{1}{l}{\textbf{LV}} & \multicolumn{1}{l}{\textbf{RV}} & \multicolumn{1}{l|}{\textbf{LVM}} & \multicolumn{1}{l}{\textbf{LV}} & \multicolumn{1}{l}{\textbf{RV}} & \multicolumn{1}{l|}{\textbf{LVM}} \\ 
			\hline
			\rowcolor{LightGreen}
			DN-Brier &  &  \phantom{x}6.7$\pm$3.1 & \phantom{x}13.5$\pm$5.9 & \phantom{x}10.2$\pm$6.9 & \phantom{x}10.7$\pm$7.7 & \phantom{x}16.7$\pm$6.8 & \phantom{x}12.3$\pm$5.8 \\
			& \textbf{e-map} &  *5.7$\pm$2.7 & *11.7$\pm$5.2 & *\phantom{x}8.3$\pm$5.9 & *\phantom{x}8.0$\pm$6.5 & *14.2$\pm$5.6 & *\phantom{x}9.7$\pm$5.0 \\
			\hdashline[1pt/2pt]
			
			\rowcolor{LightCyan}
			DN-Brier+MC &  &  \phantom{x}6.9$\pm$3.3 & \phantom{x}13.1$\pm$5.2 & \phantom{xx}9.9$\pm$5.9 & \phantom{xx}9.9$\pm$5.7 & \phantom{x}15.0$\pm$6.1 & \phantom{x}12.0$\pm$5.2 \\
			& \textbf{b-map} &  *5.5$\pm$2.6 & *10.6$\pm$5.1 & *\phantom{x}7.4$\pm$4.2 & *\phantom{x}7.5$\pm$6.0 & *12.6$\pm$5.6 & *\phantom{x}8.8$\pm$4.0 \\
			\hdashline[5pt/5pt]
			
			\rowcolor{LightGreen}
			DN-soft-Dice &  &  \phantom{x}7.1$\pm$3.5 & \phantom{x}14.8$\pm$6.8 & \phantom{x}11.0$\pm$6.6 & \phantom{x}10.2$\pm$5.6 & \phantom{x}17.7$\pm$7.8 & \phantom{x}12.9$\pm$6.2 \\
			& \textbf{e-map} &  *5.6$\pm$2.8 & *12.6$\pm$5.5 & *\phantom{x}8.6$\pm$4.6 & *\phantom{x}8.0$\pm$5.0 & *14.6$\pm$5.9 & *\phantom{x}9.6$\pm$4.5 \\
			\hdashline[1pt/2pt]
			
			\rowcolor{LightCyan}
			DN-soft-Dice+MC &  &  \phantom{x}7.7$\pm$3.9 & \phantom{x}14.4$\pm$6.0 & \phantom{x}10.5$\pm$4.9 & \phantom{x}10.1$\pm$5.3 & \phantom{x}17.2$\pm$8.0 & \phantom{x}12.5$\pm$5.3 \\
			& \textbf{b-map} &  *6.3$\pm$3.4 & *11.5$\pm$4.0 & *\phantom{x}8.6$\pm$4.8 & *\phantom{x}7.8$\pm$4.6 & *13.6$\pm$4.9 & *\phantom{x}9.6$\pm$4.7 \\
			\hdashline[5pt/5pt]
			
			\rowcolor{LightGreen}
			DRN-CE &  &  \phantom{x}5.5$\pm$2.6 & \phantom{x}11.7$\pm$5.4 & \phantom{xx}8.2$\pm$6.2 & \phantom{xx}9.1$\pm$6.4 & \phantom{x}13.7$\pm$5.6 & \phantom{xx}8.9$\pm$5.3 \\
			& \textbf{e-map} &  *4.5$\pm$1.9 & *\phantom{x}9.0$\pm$4.5 & *\phantom{x}6.3$\pm$4.1 & *\phantom{x}6.2$\pm$4.4 & *11.1$\pm$5.3 & \textbf{\textcolor{red}{*\phantom{x}6.7}}$\pm$4.2 \\
			\hdashline[1pt/2pt]
			
			\rowcolor{LightCyan}
			DRN-CE+MC &  &  \phantom{x}5.6$\pm$2.6 & \phantom{x}11.9$\pm$5.5 & \phantom{xx}8.0$\pm$5.9 & \phantom{xx}8.7$\pm$5.5 & \phantom{x}13.5$\pm$5.9 & \phantom{xx}\textbf{8.5}$\pm$4.5 \\		 
			& \textbf{b-map} &  \textbf{\textcolor{red}{*4.2}}$\pm$1.6 & \textbf{\textcolor{red}{*\phantom{x}8.1}}$\pm$3.7 & \textbf{\textcolor{red}{*\phantom{x}6.1}}$\pm$4.2 & \textbf{\textcolor{red}{*\phantom{x}5.4}}$\pm$3.6 & \textbf{\textcolor{red}{*10.1}}$\pm$5.5 & *\phantom{x}6.8$\pm$3.8 \\
			\hdashline[5pt/5pt]

			\rowcolor{LightGreen}
			DRN-soft-Dice &  &  \phantom{x}\textbf{5.5}$\pm$2.8 & \phantom{x}11.9$\pm$6.1 & \phantom{xx}\textbf{7.7}$\pm$5.9 & \phantom{xx}8.5$\pm$5.0 & \phantom{x}13.5$\pm$5.5 & \phantom{xx}8.9$\pm$5.1 \\
			& \textbf{e-map} &  *4.6$\pm$2.2 & *\phantom{x}9.4$\pm$4.5 & \phantom{xx}6.7$\pm$4.7 & *\phantom{x}6.7$\pm$4.4 & *11.6$\pm$5.4 & *\phantom{x}7.0$\pm$3.3 \\
			\hdashline[1pt/2pt]

			\rowcolor{LightCyan}
			DRN-soft-Dice+MC &  &  \phantom{x}5.7$\pm$3.2 & \phantom{x}\textbf{11.5}$\pm$5.1 & \phantom{xx}8.0$\pm$5.5 & \phantom{xx}\textbf{8.3}$\pm$4.5 & \phantom{x}\textbf{13.3}$\pm$5.1 & \phantom{xx}8.9$\pm$5.1 \\
			& \textbf{b-map} &  *4.5$\pm$2.2 & *\phantom{x}9.3$\pm$4.5 & *\phantom{x}6.3$\pm$4.0 & *\phantom{x}6.2$\pm$4.1 & *10.4$\pm$5.0 & *\phantom{x}7.0$\pm$3.4 \\
			\hdashline[5pt/5pt]
			
			\rowcolor{LightGreen}
			U-net-CE &  &  \phantom{x}6.4$\pm$4.3 & \phantom{x}15.7$\pm$8.6 & \phantom{xx}9.0$\pm$6.0 & \phantom{xx}9.7$\pm$5.3 & \phantom{x}17.0$\pm$7.7 & \phantom{x}12.7$\pm$8.2 \\
			& \textbf{e-map} &  *4.9$\pm$3.9 & *12.2$\pm$8.1 & *\phantom{x}7.1$\pm$5.6 & *\phantom{x}6.1$\pm$3.2 & *12.6$\pm$6.5 & *\phantom{x}8.4$\pm$6.3 \\
			\hdashline[1pt/2pt]

			\rowcolor{LightCyan}
			U-net-CE+MC &  &  \phantom{x}6.2$\pm$4.2 & \phantom{x}15.3$\pm$8.4 & \phantom{xx}8.8$\pm$5.8 & \phantom{xx}9.2$\pm$5.0 & \phantom{x}16.5$\pm$7.6 & \phantom{x}12.0$\pm$8.0 \\
			& \textbf{b-map} &  *4.3$\pm$1.6 & *\phantom{x}9.9$\pm$6.6 & *\phantom{x}6.7$\pm$4.8 & *\phantom{x}5.4$\pm$2.8 & *10.3$\pm$4.7 & *\phantom{x}7.6$\pm$6.2 \\
			
			\hdashline[5pt/5pt]

			\rowcolor{LightGreen}
			U-net-soft-Dice &  &  \phantom{x}6.1$\pm$3.9 & \phantom{x}14.1$\pm$7.6 & \phantom{x}10.6$\pm$8.4 & \phantom{xx}9.2$\pm$7.1 & \phantom{x}16.3$\pm$7.5 & \phantom{x}12.6$\pm$9.6 \\
			& \textbf{e-map} &  *4.6$\pm$2.3 & *11.3$\pm$7.2 & *\phantom{x}7.5$\pm$5.5 & *\phantom{x}7.3$\pm$6.5 & *13.7$\pm$7.6 & *\phantom{x}9.8$\pm$8.0 \\
			\hdashline[1pt/2pt]
			
			\rowcolor{LightCyan}
			U-net-soft-Dice+MC &  &  \phantom{x}6.2$\pm$3.9 & \phantom{x}14.1$\pm$7.7 & \phantom{x}10.5$\pm$8.7 & \phantom{xx}9.0$\pm$7.0 & \phantom{x}15.8$\pm$7.5 & \phantom{x}12.1$\pm$9.2 \\
			& \textbf{b-map} &  *4.5$\pm$2.1 & *10.4$\pm$7.2 & *\phantom{x}7.6$\pm$7.0 & *\phantom{x}7.3$\pm$6.9 & *12.9$\pm$6.6 & *\phantom{x}9.8$\pm$8.4 \\
			\hdashline
			Isensee et al. &  & \phantom{x}7.1 &  \phantom{x}14.3 & \phantom{xx}8.9 &  \phantom{xx}9.8 &  \phantom{x}16.3 & \phantom{x}10.4 \\
			\hline
			
		\end{tabular}
		\label{table_seg_perf_hd}
	}   
	
\end{table*}

\section{Evaluation}\label{evaluation}

Automatic segmentation performance, as well as performance after simulating the correction of detected segmentation failures and after manual expert correction was evaluated. For this, the \num{3}D Dice-coefficient (DC) and \num{3}D Hausdorff distance (HD) between manual and (corrected) automatic segmentation were computed. Furthermore, the following clinical metrics were computed for manual and (corrected) automatic segmentation: left ventricle end-diastolic volume (EDV); left ventricle ejection fraction (EF); right ventricle EDV; right ventricle ejection fraction; and left ventricle myocardial mass. Following Bernard et al.\cite{bernard2018deep} for each of the clinical metrics three performance indices were computed using the measurements based on manual and (corrected) automatic segmentation: Pearson correlation coefficient; mean difference (bias and standard deviation); and mean absolute error (MAE).

To evaluate detection performance of the automatic method precision-recall curves of identification of slices that require correction were computed. A slice is considered positive in case it consists of at least one image region with a segmentation failure. To achieve accurate segmentation in clinic, identification of slices that contain segmentation failures might ease manual correction of automatic segmentations in daily practice. To further evaluate detection performance detection rate of segmentation failures was assessed on a voxel level. More specific, sensitivity against the number of false positive regions was evaluated because manual correction is presumed to be performed at this level.

Finally, after simulation and manual correction of the automatically detected segmentation failures, segmentation was re-evaluated and significance of the difference between the DCs, HDs and clinical metrics was tested with a Mann–Whitney U test.


\section{Experiments}

To use stratified four-fold cross-validation the dataset was split into training (75\%) and test (25\%) set. The splitting was done on a patient level, so there was no overlap in patient data between training and test sets. Furthermore, patients were randomly chosen from each of the five patient groups w.r.t. disease. Each patient has one volume for ED and ES time points, respectively.  

\subsection{Training segmentation networks} \label{training_segmentation}

DRN and U-net were trained with a patch size of \num{128}$\times$\num{128} voxels which is a multiple of their output stride of the contracting path. In the training of the dilated CNN (DN) images with \num{151}$\times$\num{151} voxel samples were used. Zero-padding to \num{281}$\times$\num{281} was performed to accommodate the \num{131}$\times$\num{131} voxel receptive field that is induced by the dilation factors. Training samples were randomly chosen from training set and augmented by \num{90} degree rotations of the images. All models were initially trained with three loss functions: soft-Dice\cite{milletari2016v} (SD); cross-entropy (CE); and Brier loss\cite{brier1950verification}. However, for the evaluation of the combined segmentation and detection approach for each model architecture the two best performing loss functions were chosen: soft-Dice for all models; cross-entropy for DRN and U-net and Brier loss for DN. For completeness, we provide the equations for all three used loss functions.

\begingroup
\small
\begin{align}
\text{soft-Dice}_{c} = \frac{\sum_{i=1}^{N} R_{c}(i) \; A_{c}(i) }{\sum_{i=1}^{N} R_{c}(i) + \sum_{i=1}^{N} A_{c}(i)}  \; ,
\end{align}
\endgroup
where $N$ denotes the number of voxels in an image, $R_{c}$ is the binary reference image for class $c$ and $A_{c}$ is the probability map for class $c$.

\begingroup
\small
\begin{align}
\begin{split}
\text{Cross-Entropy}_{c} &= - \; \sum_{i=1}^{N} t_{ic} \; \log \; p(y_i=c|x_i) \; , \\& \text{ where } t_{ic} = 1 \text{ if } y_{i}=c, \text{ and \num{0} otherwise.}
\end{split}
\end{align}
\endgroup

\begingroup
\small
\begin{align}
\begin{split}
\text{Brier}_{c} &= \sum_{i=1}^{N}  \big(t_{ic} - p(y_i=c|x_{i}) \big)^2 \; , \\ &\text{ where } t_{ic} = 1 \text{ if } y_{i}=c, \text{ and \num{0} otherwise.}
\end{split}
\end{align}
\endgroup

where $N$ denotes the number of voxels in an image and $p$ denotes the probability for a specific voxel $x_i$ with corresponding reference label $y_i$ for class $c$.

Choosing Brier loss to train the DN model instead of CE was motivated by our preliminary work which showed that segmentation performance of DN model was best when trained with Brier loss\cite{sander2019towards}.

All models were trained for 100,000 iterations. DRN and U-net were trained with a learning rate of \num{0.001} which decayed with a factor of \num{0.1} after every 25,000 steps. Training DN used the snapshot ensemble technique~\cite{huang2017snapshot}, where after every 10,000 iterations the learning rate was reset to its original value of \num{0.02}.

All three segmentation networks were trained using mini-batch stochastic gradient descent using a batch size of \num{16}. Network parameters were optimized using the Adam optimizer \cite{kingmadp}. Furthermore, models were regularized with weight decay to increase generalization performance. 

\subsection{Training detection network}\label{label_training_detection}

To train the detection model a subset of the errors performed by the segmentation model is used. Segmentation errors that presumably are within the range of inter-observer variability and therefore do not inevitably require correction (tolerated errors) are excluded from the set of errors that need to be detected and corrected (segmentation failures). To distinguish between tolerated errors and the set of segmentation failures $\mathcal{S}_I$ the Euclidean distance of an incorrectly segmented voxel to the boundary of the reference target structure is used. For each anatomical structure a \num{2}D distance transform map is computed that provides for each voxel the distance to the anatomical structure boundary. To differentiate between tolerated errors and the set of segmentation failures $\mathcal{S}_I$ an acceptable tolerance threshold is applied. A more rigorous threshold is used for errors located inside compared to outside of the anatomical structure because automatic segmentation methods have a tendency to undersegment cardiac structures in CMRI. Hence, in all experiments the acceptable tolerance threshold was set to three voxels (equivalent to on average \SI{4.65}{\milli\meter}) and two voxels (equivalent to on average \SI{3.1}{\milli\meter}) for segmentation errors located outside and inside the target structure. Furthermore, a segmentation error only belongs to $\mathcal{S}_I$ if it is part of a \num{2}D \num{4}-connected cluster of minimum size \num{10} voxels. This value was found in preliminary experiments by evaluating values $\{1, 5, 10, 15, 20\}$. However, for apical slices all segmentation errors are included in $\mathcal{S}_I$ regardless of fulfilling the minimum size requirement because in these slices anatomical structures are relatively small and manual segmentation is prone to large inter-observer variability~\cite{bernard2018deep}. Finally, segmentation errors located in slices above the base or below the apex are always included in the set of segmentation failures.

Using the set $\mathcal{S}_I$ a binary label $t_j$ is assigned to each patch $P_j^{(I)}$ indicating whether $P_j^{(I)}$ contains at least one voxel belonging to set $\mathcal{S}_I$ where $j \in \{1 \dots M \}$ and $M$ denotes the number of patches in a slice $I$.  

The detection network is trained by minimizing a weighted binary cross-entropy loss:

\begingroup
\small
\begin{equation} \label{eq_detection_loss}
\mathcal{L}_{DT} = - \sum_{j \in P^{(I)}} w_{pos} \; t_j \log p_j + (1 - t_j) \log (1 - p_j)  \; ,
\end{equation}
\endgroup

where $w_{pos}$ represents a scalar weight, $t_j$ denotes the binary reference label and $p_j$ is the softmax probability indicating whether a particular image region $P_j^{(I)}$ contains at least one segmentation failure. The average percentage of regions in a patient volume containing segmentation failures ranges from \num{1.5} to \num{3} percent depending on the segmentation architecture and loss function used to train the segmentation model. To train a detection network $w_{pos}$ was set to the ratio between the average percentage of negative samples divided by the average percentage of positive samples.

Each fold was trained using spatial uncertainty maps and automatic segmentation masks generated while training the segmentation networks. Hence, there was no overlap in patient data between training and test set across segmentation and detection tasks. In total \num{12} detection models were trained and evaluated resulting from the different combination of \num{3} model architectures (DRN, DN and U-net), \num{2} loss functions (DRN and U-net with CE and soft-Dice, DN with Brier and soft-Dice) and \num{2} uncertainty maps (e-maps, b-maps).

\begin{table*}
	\caption{Segmentation performance of different combination of model architectures, loss functions and evaluation modes (without or with MC dropout (MC) enabled during testing) in terms of clinical metrics: left ventricle (LV) end-diastolic volume (EDV); LV ejection fraction (EF); right ventricle (RV) EDV; RV ejection fraction; and LV myocardial mass. Quantitative results compare clinical metrics based on reference segmentations with 1) automatic segmentations and 2) simulated manual correction of automatic segmentations using spatial uncertainty maps. $\rho$ denotes the Pearson correlation coefficient, \textit{bias} denotes the mean difference between the two measurements (mean $\pm$ standard deviation) and \textit{MAE} denotes the mean absolute error between the two measurements. Each combination comprises a block of two rows. A row in which column \textit{Uncertainty map for detection} indicates e- or b-map shows results for the combined segmentation and detection approach. Numbers accentuated in black/bold are ranked first in the segmentation only task. Numbers in red indicate statistical significant at \num{5}\% level w.r.t. the segmentation-only approach for the specific clinical metric. Best viewed in color.}
	\label{table_cardiac_function_indices}
	\tiny
	\centering
	\begin{tabular}{| C{1.6cm} | C{1.cm} | C{0.3cm} c C{0.3cm} | C{0.2cm} C{0.8cm} C{0.3cm} | C{0.3cm} C{0.8cm} C{0.3cm} | C{0.3cm} C{0.8cm} C{0.3cm} | C{0.3cm} C{0.8cm} C{0.3cm} |}
		\hline
		& \multicolumn{1}{c|}{\thead{\textbf{Uncertainty} \\ \textbf{map for} \\ \textbf{detection}}}  & \multicolumn{3}{c}{\textbf{LV$_{EDV}$}} &  \multicolumn{3}{c}{\textbf{LV$_{EF}$}} & \multicolumn{3}{c}{\textbf{RV$_{EDV}$}} &  \multicolumn{3}{c}{\textbf{RV$_{EF}$}} & \multicolumn{3}{c|}{\textbf{LVM$_{Mass}$}}  \\
		\textbf{Method} &  & \textbf{$\rho$} & \multicolumn{1}{l}{\textbf{bias$\pm\sigma$}} & \textbf{MAE} & \textbf{$\rho$} & \textbf{bias$\pm \sigma$} & \textbf{MAE} & \textbf{$\rho$} & \textbf{bias$\pm \sigma$} & \textbf{MAE} & \textbf{$\rho$} & \textbf{bias$\pm \sigma$} & \textbf{MAE} & \multicolumn{1}{c}{\textbf{$\rho$}} & \textbf{bias$\pm \sigma$} & \textbf{MAE} \\
		\hline
		\rowcolor{LightGreen}
		DN-Brier &  & 0.997 & \phantom{x}\textbf{0.0$\pm$6.1} & 4.5 & 0.892 & \phantom{x}2.2$\pm$\phantom{x}9.2 & 4.2 & 0.977 & \textbf{-0.2$\pm$11.8} & \phantom{x}8.5 & 0.834 & 5.3$\pm$10.3 & \phantom{x}8.5 & 0.984 & -2.7$\pm$\phantom{x}9.0 & \phantom{x}\textbf{7.0} \\
		& e-map  & 0.997 & \phantom{x}0.0$\pm$5.5 & 4.0 & 0.982 & \phantom{x}0.1$\pm$\phantom{x}3.8 & 2.2 & 0.992 & \phantom{x}0.0$\pm$\phantom{x}6.9 & \phantom{x}5.2 & 0.955 & 1.9$\pm$\phantom{x}5.5 & \phantom{x}4.1 & 0.986 & -2.1$\pm$\phantom{x}8.4 & \phantom{x}6.6 \\
		
		\hdashline[1pt/2pt]
		\rowcolor{LightCyan}
		DN-Brier+MC &  & 0.997 & \phantom{x}1.6$\pm$6.0 & 4.4 & 0.921 & \phantom{x}1.1$\pm$\phantom{x}7.9 & 3.9 & 0.975 & \phantom{x}6.7$\pm$12.4 & \phantom{x}9.6 & 0.854 & 3.5$\pm$\phantom{x}9.9 & \phantom{x}7.7 & 0.984 & \phantom{x}0.7$\pm$\phantom{x}9.2 & \phantom{x}7.1 \\
		& b-map  & 0.998 & \phantom{x}1.0$\pm$5.3 & 3.9 & 0.991 & \phantom{x}0.0$\pm$\phantom{x}2.7 & 1.9 & 0.993 & \phantom{x}3.2$\pm$\phantom{x}6.7 & \phantom{x}5.7 & 0.975 & 0.8$\pm$\phantom{x}4.0 & \phantom{x}3.0 & 0.987 & \phantom{x}0.1$\pm$\phantom{x}8.3 & \phantom{x}6.5 \\
		\hdashline[1pt/2pt]
		\rowcolor{LightGreen}
		DN-soft-Dice &  & 0.996 & \phantom{x}1.2$\pm$6.5 & 4.9 & 0.918 & \phantom{x}1.5$\pm$\phantom{x}8.0 & 3.9 & 0.972 & \phantom{x}\textbf{0.2$\pm$13.0} & \phantom{x}9.6 & 0.802 & 7.2$\pm$11.3 & 10.2 & 0.982 & -4.5$\pm$\phantom{x}9.6 & \phantom{x}8.5 \\
		& e-map  & 0.997 & \phantom{x}1.0$\pm$5.5 & 4.2 & 0.989 & \phantom{x}0.2$\pm$\phantom{x}3.0 & 2.2 & 0.990 & \phantom{x}0.2$\pm$\phantom{x}7.6 & \phantom{x}5.9 & \textcolor{red}{0.940} & \textcolor{red}{3.3$\pm$\phantom{x}6.2} & \textcolor{red}{\phantom{x}5.2} & 0.983 & -4.3$\pm$\phantom{x}9.3 & \phantom{x}8.2 \\
		
		\hdashline[1pt/2pt]
		\rowcolor{LightCyan}
		DN-soft-Dice+MC &  & 0.996 & \phantom{x}3.2$\pm$7.1 & 5.6 & 0.958 & \phantom{x}0.4$\pm$\phantom{x}5.7 & 3.6 & 0.964 & \phantom{x}8.1$\pm$14.9 & 12.3 & 0.827 & 4.8$\pm$11.0 & \phantom{x}8.9 & 0.978 & -0.7$\pm$10.7 & \phantom{x}8.3 \\
		& b-map  & 0.997 & \phantom{x}2.2$\pm$5.6 & 4.4 & 0.988 & -0.2$\pm$\phantom{x}3.1 & 2.2 & 0.990 & \phantom{x}4.0$\pm$\phantom{x}7.7 & \phantom{x}7.0 & 0.959 & 1.8$\pm$\phantom{x}5.1 & \phantom{x}4.1 & 0.982 & -1.4$\pm$\phantom{x}9.5 & \phantom{x}7.6 \\
		
		\hdashline[5pt/5pt]
		\rowcolor{LightGreen}
		DRN-CE &  & 0.997 & -0.2$\pm$5.5 & 4.1 & 0.968 & \phantom{x}1.2$\pm$\phantom{x}5.0 & 3.5 & 0.976 & \phantom{x}1.5$\pm$12.1 & \phantom{x}8.5 & 0.870 & 1.3$\pm$\phantom{x}9.2 & \phantom{x}6.9 & 0.980 & \phantom{x}\textbf{0.6$\pm$10.2} & \phantom{x}7.8 \\
		& e-map  & 0.998 & \phantom{x}0.2$\pm$4.5 & 3.5 & 0.992 & \phantom{x}0.2$\pm$\phantom{x}2.5 & 1.9 & 0.988 & \phantom{x}1.4$\pm$\phantom{x}8.5 & \phantom{x}6.2 & 0.952 & 0.8$\pm$\phantom{x}5.6 & \phantom{x}4.2 & 0.985 & \phantom{x}0.4$\pm$\phantom{x}8.7 & \phantom{x}6.8 \\
		
		\hdashline[1pt/2pt]
		\rowcolor{LightCyan}
		DRN-CE+MC &  & \textbf{0.998} & \phantom{x}1.0$\pm$4.9 & \textbf{3.9} & 0.972 & \phantom{x}0.8$\pm$\phantom{x}4.6 & 3.1 & 0.973 & \phantom{x}4.8$\pm$12.8 & \phantom{x}9.4 & 0.876 & \textbf{0.4$\pm$\phantom{x}9.1} & \phantom{x}\textbf{6.6} & 0.981 & \phantom{x}1.9$\pm$\phantom{x}9.9 & \phantom{x}7.6 \\
		& b-map  & 0.998 & \phantom{x}0.7$\pm$4.6 & 3.6 & 0.992 & -0.1$\pm$\phantom{x}2.5 & 1.8 & 0.992 & \phantom{x}2.9$\pm$\phantom{x}6.9 & \phantom{x}5.7 & 0.967 & 0.6$\pm$\phantom{x}4.6 & \phantom{x}3.4 & 0.987 & \phantom{x}1.2$\pm$\phantom{x}8.3 & \phantom{x}6.6 \\
		\hdashline[1pt/2pt]
		
		\rowcolor{LightGreen}
		DRN-soft-Dice &  & \textbf{0.998} & \phantom{x}0.8$\pm$5.1 & 4.0 & 0.976 & \phantom{x}0.2$\pm$\phantom{x}4.4 & 3.0 & 0.980 & \phantom{x}\textbf{0.2$\pm$11.0} & \phantom{x}\textbf{7.5} & \textbf{0.882} & 3.1$\pm$\phantom{x}8.7 & \phantom{x}6.8 & 0.984 & -3.5$\pm$\phantom{x}9.1 & \phantom{x}7.5 \\
		& e-map  & 0.998 & \phantom{x}0.7$\pm$4.4 & 3.5 & 0.987 & -0.1$\pm$\phantom{x}3.1 & 2.2 & 0.987 & \phantom{x}0.1$\pm$\phantom{x}9.1 & \phantom{x}6.4 & 0.938 & 1.9$\pm$\phantom{x}6.3 & \phantom{x}4.9 & 0.986 & -3.5$\pm$\phantom{x}8.7 & \phantom{x}7.1 \\
		
		\hdashline[1pt/2pt]
		\rowcolor{LightCyan}
		DRN-soft-Dice+MC &   & \textbf{0.998} & \phantom{x}1.8$\pm$5.1 & \textbf{3.9} & \textbf{0.979} & -0.3$\pm$\phantom{x}4.1 & 2.9 & 0.977 & \phantom{x}3.5$\pm$11.7 & \phantom{x}8.1 & 0.868 & 1.7$\pm$\phantom{x}9.5 & \phantom{x}6.8 & 0.983 & -1.4$\pm$\phantom{x}9.5 & \phantom{x}7.4 \\
		&  b-map  & 0.998 & \phantom{x}1.7$\pm$4.7 & 3.7 & 0.990 & -0.2$\pm$\phantom{x}2.9 & 2.1 & 0.989 & \phantom{x}2.3$\pm$\phantom{x}8.1 & \phantom{x}5.8 & 0.959 & 0.8$\pm$\phantom{x}5.2 &\phantom{x}3.8 & 0.986 & -1.3$\pm$\phantom{x}8.5 & \phantom{x}6.8 \\
		
		\hdashline[5pt/5pt]
		\rowcolor{LightGreen}
		U-net-CE &  & 0.995 & -4.7$\pm$7.2 & 6.1 & 0.954 & \phantom{x}4.1$\pm$\phantom{x}6.0 & 5.1 & 0.963 & -7.6$\pm$15.2 & 12.1 & 0.870 & 5.6$\pm$\phantom{x}9.0 & \phantom{x}8.1 & 0.971 & -8.5$\pm$12.2 & 11.5 \\
		& e-map  & 0.998 & -3.2$\pm$4.8 & 4.4 & 0.992 & \phantom{x}1.7$\pm$\phantom{x}2.6 & 2.4 & 0.987 & -4.1$\pm$\phantom{x}9.1 & \phantom{x}6.7 & 0.957 & 2.6$\pm$\phantom{x}5.2 & \phantom{x}4.1 & 0.983 & -5.7$\pm$\phantom{x}9.3 & \phantom{x}8.2 \\
		
		\hdashline[1pt/2pt]
		\rowcolor{LightCyan}
		U-net-CE+MC &   & 0.995 & -4.3$\pm$7.2 & 5.9 & 0.958 & \phantom{x}3.8$\pm$\phantom{x}5.8 & 4.9 & 0.968 & -4.8$\pm$14.1 & 10.7 & 0.867 & 5.0$\pm$\phantom{x}9.1 & \phantom{x}7.9 & 0.972 & -8.1$\pm$12.0 & 11.1 \\
		& b-map  & 0.997 & -3.5$\pm$5.5 & 4.9 & 0.990 & \phantom{x}1.6$\pm$\phantom{x}2.9 & 2.6 & 0.992 & -1.8$\pm$\phantom{x}7.0 & \phantom{x}4.9 & 0.974 & 1.6$\pm$\phantom{x}4.1 & \phantom{x}3.3 & 0.981 & -6.8$\pm$10.0 & \phantom{x}9.4 \\
		\hdashline[1pt/2pt]
		
		\rowcolor{LightGreen}
		U-net-soft-Dice &  & 0.997 & -2.0$\pm$6.0 & 4.5 & 0.853 & \phantom{x}3.6$\pm$10.9 & 5.0 & 0.968 & -1.0$\pm$14.1 & 10.0 & 0.782 & 4.8$\pm$11.6 & \phantom{x}9.0 & \textbf{0.985} & -7.7$\pm$\phantom{x}8.8 & \phantom{x}9.2 \\
		& e-map  & 0.997 & -1.7$\pm$5.3 & 4.1 & 0.969 & \phantom{x}1.9$\pm$\phantom{x}4.9 & 3.3 & 0.981 & -0.1$\pm$10.9 & \phantom{x}7.5 & 0.919 & 3.3$\pm$\phantom{x}7.0 & \phantom{x}5.9 & 0.984 & -6.6$\pm$\phantom{x}9.0 & \phantom{x}8.7 \\
		
		\hdashline[1pt/2pt]
		\rowcolor{LightCyan}
		U-net-soft-Dice+MC &  & 0.997 & -1.8$\pm$5.9 & 4.4 & 0.941 & \phantom{x}3.0$\pm$\phantom{x}6.7 & 4.4 & 0.969 & \phantom{x}0.6$\pm$13.9 & \phantom{x}9.8 & 0.792 & 4.4$\pm$11.3 & \phantom{x}8.7 & \textbf{0.985} & -7.2$\pm$\phantom{x}8.9 & \phantom{x}8.9 \\
		& b-map  & 0.997 & -1.5$\pm$5.3 & 4.1 & 0.979 & \phantom{x}1.1$\pm$\phantom{x}4.1 & 2.9 & 0.985 & \phantom{x}1.2$\pm$\phantom{x}9.4 & \phantom{x}6.5 & 0.939 & 2.9$\pm$\phantom{x}6.2 & \phantom{x}4.9 & 0.984 & -5.9$\pm$\phantom{x}9.0 & \phantom{x}8.5 \\
		\hline
	\end{tabular}
\end{table*}

The patches used to train the network were selected randomly (\nicefrac{2}{3}), or were forced (\nicefrac{1}{3}) to contain at least one segmentation failure by randomly selecting a scan containing segmentation failure, followed by random sampling of a patch containing at least one segmentation failure. During training the patch size was fixed to \num{80}$\times$\num{80} voxels. To reduce the number of background voxels during testing, inputs were cropped based on a minimal enclosing, rectangular bounding box that was placed around the automatic segmentation mask. Inputs always had a minimum size of \num{80}$\times$\num{80} voxels or were forced to a multiple of the output grid spacing of eight voxels in both direction required by the patch-based detection network. The patches of size \num{8}$\times$\num{8} voxels did not overlap. In cases where the automatic segmentation mask only contains background voxels (scans above the base or below apex of the heart) input scans were center-cropped to a size of \num{80}$\times$\num{80} voxels. 

Models were trained for 20,000 iterations using mini-batch stochastic gradient descent with batch-size \num{32} and Adam as optimizer\cite{kingmadp}. Learning rate was set to \num{0.0001} and decayed with a factor of \num{0.1} after \num{10,000} steps. Furthermore, dropout percentage was set to \num{0.5} and weight decay was applied to increase generalization performance.

\begin{figure*}[t]
	\center
	\subfloat[]{\includegraphics[width=3.4in, height=1.7in]{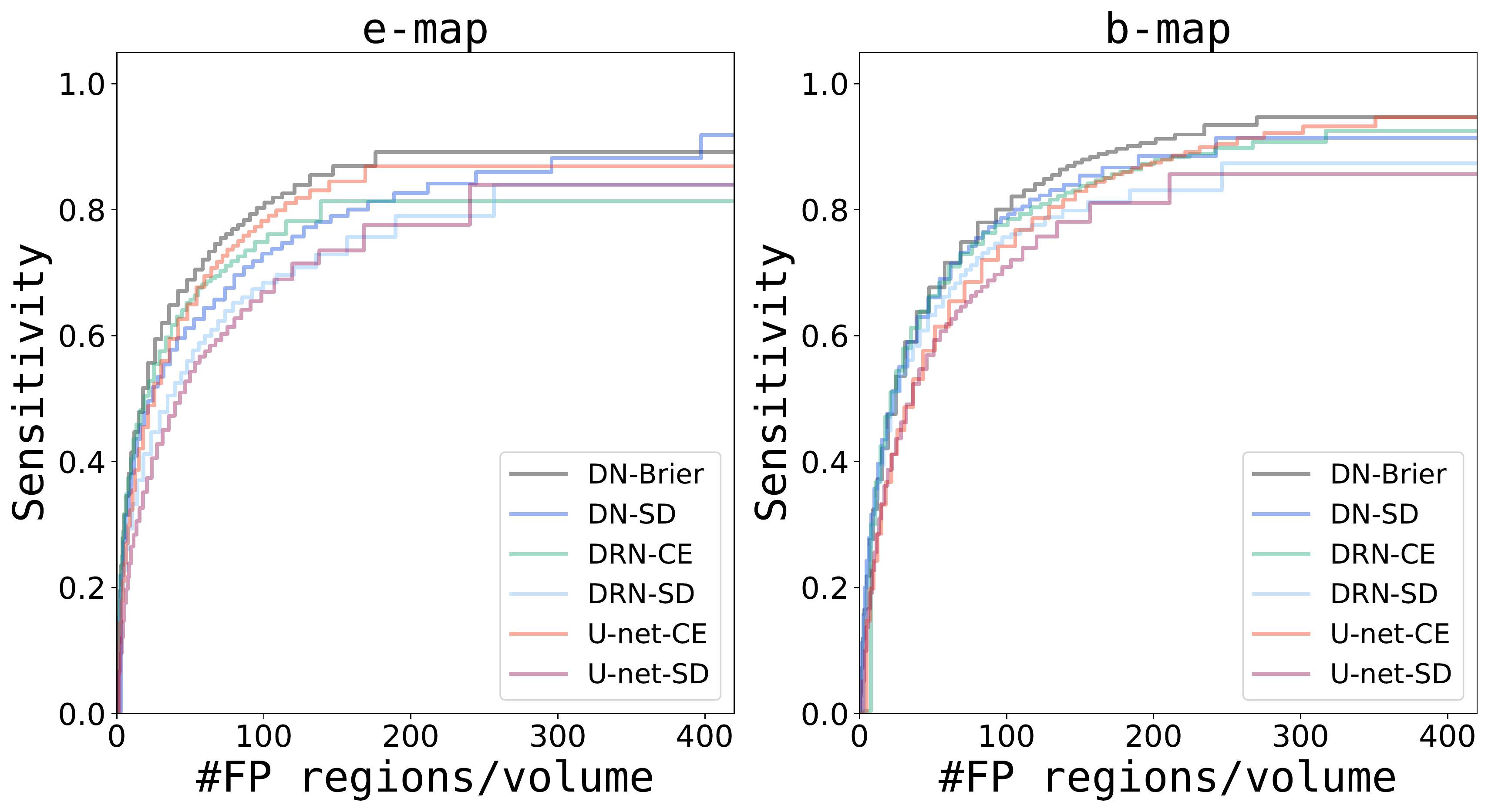}%
		\label{fig_froc_voxel_detection}}
	\subfloat[]{\includegraphics[width=3.4in, height=1.7in]{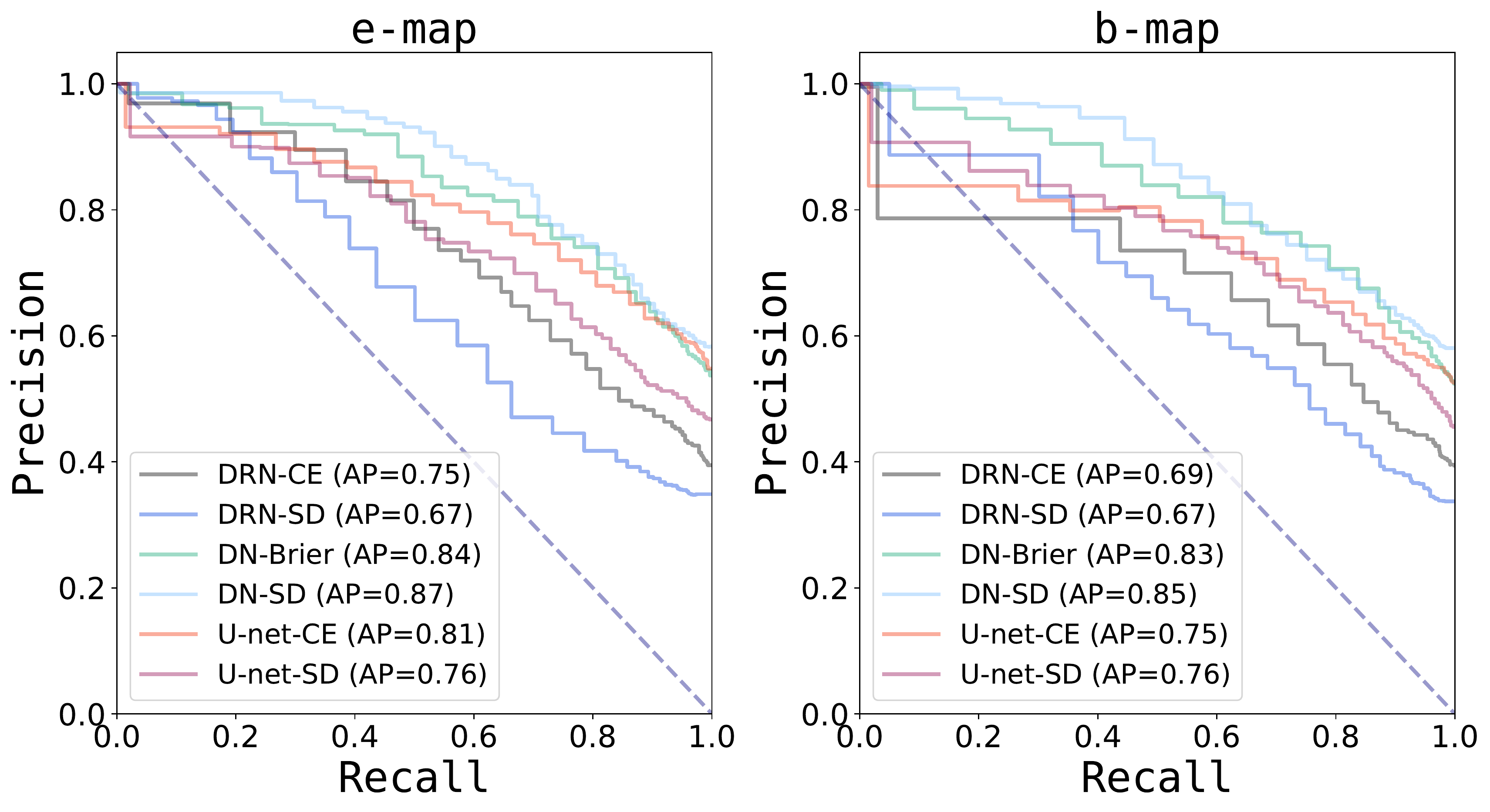}%
		\label{fig_prec_rec_slice_detection}}
	
	\caption{Detection performance of segmentation failures generated by different combination of segmentation architectures and loss functions. (a) Sensitivity for detection of segmentation failures on voxel level (y-axis) as a function of number of false positive image regions (x-axis). (b) Precision-recall curve for detection of slices containing segmentation failures (where AP denotes average precision). Results are split between entropy and Bayesian uncertainty maps. Each figure contains a curve for the six possible combination of models (three) and loss functions (two). SD denotes soft-Dice and CE cross-entropy, respectively.}
	\label{fig_dt_perf_all_models}
\end{figure*}

\subsection{Segmentation using correction of the detected segmentation failures}

To investigate whether correction of detected segmentation failures increases segmentation performance two scenarios were performed. In the first scenario manual correction of the detected failures by an expert was simulated for all images at ED and ES time points of the ACDC dataset. For this purpose, in image regions that were detected to contain segmentation failure predicted labels were replaced with reference labels. In the second scenario manual correction of the detected failures was performed by an expert in a random subset of \num{50} patients of the ACDC dataset. The expert was shown CMRI slices for ED and ES time points together with corresponding automatic segmentation masks for the RV, LV and LV myocardium. Image regions detected to contain segmentation failures were indicated in slices and the expert was only allowed to change the automatic segmentations in these indicated regions. Annotation was performed following the protocol described in\cite{bernard2018deep}. Furthermore, expert was able to navigate through all CMRI slices of the corresponding ED and ES volumes.

\section{Results}

In this section we first present results for the segmentation-only task followed by description of the combined segmentation and detection results. 

\subsection{Segmentation-only approach} \label{results_seg_only}

Table~\ref{table_overall_segmentation_performance} lists quantitative results for segmentation-only and combined segmentation and detection approach in terms of Dice coefficient and Hausdorff distance. These results show that DRN and U-net achieve similar Dice coefficients and outperformed the DN network for all anatomical structures at end-systole. Differences in the achieved Hausdorff distances among the methods are present for all anatomical structures and for both time points. The DRN model achieved the highest and the DN network the lowest Hausdorff distance.

Table~\ref{table_cardiac_function_indices} lists results of the evaluation in terms of clinical metrics. These results reveal noticeable differences between models for ejection fraction (EF) of left and right ventricle, respectively. We can observe that U-net trained with the soft-Dice and the Dilated Network (DN) trained with Brier or soft-Dice loss achieved considerable lower accuracy for LV and RV ejection fraction compared to DRN. Overall, the DRN model achieved highest performance for all clinical metrics.

\noindent \textbf{Effect of model architecture on segmentation}: Although quantitative differences between models are small, qualitative evaluation discloses that automatic segmentations differ substantially between the models. Figure~\ref{fig_seg_qualitative_results} shows that especially in regions where the models perform poorly (apical and basal slices) the DN model more often produced anatomically implausible segmentations compared to the DRN and U-net. This seems to be correlated with the performance differences in Hausdorff distance.

\noindent \textbf{Effect of loss function on segmentation}: The results indicate that the choice of loss function only slightly affects the segmentation performance. DRN and U-net perform marginally better when trained with soft-Dice compared to cross-entropy whereas DN performs better when trained with Brier loss than with soft-Dice. For DN this is most pronounced for the RV at ES.

A considerable effect of the loss function on the accuracy of the LV and RV ejection fraction can be observed for the U-net model. On both metrics U-net achieved the lowest accuracy of all models when trained with the soft-Dice loss.

\noindent \textbf{Effect of MC dropout on segmentation}: The results show that enabling MC-dropout during testing seems to result in slightly improved HD while it does not affect DC. 
\begin{table}
	\caption{Average precision and percentage of slices with segmentation failures generated by Dilated Network (DN), Dilated Residual Network (DRN) and U-net when trained with soft-Dice (SD), CE or Brier loss. Per patient, average precision of detected slices with failure using e- or b-maps (\num{2}$^{nd}$ and \num{3}$^{rd}$ columns). Per patient, average percentage of slices containing segmentation failures (reference for detection task) (\num{4}$^{th}$ and \num{5}$^{th}$ columns).}
	\label{table_evaluation_slice_detection}
	\begin{tabular}{l  C{1.4cm}  C{1.4cm} C{1.4cm} C{1.4cm}  }
		\textbf{Model} & \multicolumn{2}{c}{\textbf{Average precision}} & \multicolumn{2}{c}{ \thead{\textbf{\% of slices} \\ \textbf{with segmentation failures} }} \\
		& e-map & b-map & e-map & b-map \\
		\hline
		DN-Brier &  84.0 & 83.0 & 53.7 & 52.4   \\
		DN-SD  &  87.0 & 85.0 & 58.3 & 58.1   \\
		\hdashline
		DRN-CE  & 75.0 & 69.0  &  39.5 & 39.4 \\
		DRN-SD  & 67.0 & 67.0 &  34.9 & 33.7\\
		\hdashline
		U-net-CE   & 81.0 & 75.0 & 54.8 & 52.5 \\
		U-net-SD   & 76.0 & 76.0 & 46.7 & 45.5 \\
		\hline
	\end{tabular}
\end{table}

\subsection{Detection of segmentation failures}

\noindent \textbf{Detection of segmentation failures on voxel level}: To evaluate detection performance of segmentation failures on voxel level Figure~\ref{fig_froc_voxel_detection} shows average voxel detection rate as a function of false positively detected regions. This was done for each combination of model architecture and loss function exploiting e- (Figure~\ref{fig_froc_voxel_detection}, left)  or b-maps (Figure~\ref{fig_froc_voxel_detection}, right). These results show that detection performance of segmentation failures depends on segmentation model architecture, loss function and uncertainty map. 

The influence of (segmentation) model architecture and loss function on detection performance is slightly stronger when e-maps were used as input for the detection task compared to b-maps. Detection rates are consistently lower when segmentation failures originate from segmentation models trained with soft-Dice loss compared to models trained with CE or Brier loss. Overall, detection rates are higher when b-maps were exploited for the detection task compared to e-maps.

\begin{table*}
	\caption{Comparing performance of segmentation-only approach (auto-only) with combined segmentation and detection approach for two scenarios: simulated correction of detected segmentation failures (auto$+$simulation); and manual correction of detected segmentation failures by an expert (auto$+$expert). Automatic segmentations were obtained from a U-net trained with cross-entropy. Evaluation was performed on a subset of \num{50} patients from the ACDC dataset. Scenarios are compared against segmentation-only approach (auto-only) in terms of (a) Dice Coefficient (b) Hausdorff Distance and (c) Clinical metrics. Results obtained from simulated manual correction represent an upper bound on the maximum achievable performance. Detection network was trained with e-maps. Number with asterisk indicates statistical significant at \num{5}\% level w.r.t. the segmentation-only approach. Best viewed in color.}
	\label{table_manual_corr_performance}
	\centering
	\small
	\subfloat[\textbf{Dice coefficient:} Mean $\pm$ standard deviation for left ventricle (LV), right ventricle (RV) and left ventricle myocardium (LVM).]{
		\begin{tabular}{| C{2.cm} | R{1.7cm} R{1.7cm} R{1.7cm} | R{1.7cm} R{1.7cm} R{1.7cm} | }
			\hline
			& \multicolumn{3}{c|}{\textbf{End-diastole}} & \multicolumn{3}{c|}{\textbf{End-systole}} \\
			\textbf{Scenario} & \multicolumn{1}{l}{\textbf{LV}} & \multicolumn{1}{l}{\textbf{RV}} & \multicolumn{1}{l|}{\textbf{LVM}} & \multicolumn{1}{l}{\textbf{LV}} & \multicolumn{1}{l}{\textbf{RV}} & \multicolumn{1}{l|}{\textbf{LVM}} \\ 
			\hline
			auto-only & \phantom{x}0.964$\pm$0.02 & \phantom{x}0.927$\pm$0.04 & \phantom{x}0.883$\pm$0.03  & \phantom{x}0.916$\pm$0.05 & \phantom{x}0.854$\pm$0.08 & \phantom{x}0.886$\pm$0.04 \\ 
			auto$+$simulation & \phantom{x}0.967$\pm$0.01 & *0.948$\pm$0.03 & *0.894$\pm$0.03  & *0.939$\pm$0.03 & *0.915$\pm$0.04 & *0.910$\pm$0.03 \\ 
			auto$+$expert & \phantom{x}0.965$\pm$0.02 & \phantom{x}0.940$\pm$0.03 & \phantom{x}0.885$\pm$0.03  & \phantom{x}0.927$\pm$0.04 & \phantom{x}0.868$\pm$0.07 & \phantom{x}0.894$\pm$0.03 \\ 
			
				\hline
			
		\end{tabular}
		\label{table_manual_seg_perf_dsc}
	}  
	
	\centering
	
	\subfloat[\textbf{Hausdorff Distance:} Mean $\pm$ standard deviation for left ventricle (LV), right ventricle (RV) and left ventricle myocardium (LVM).]{
		\begin{tabular}{| C{2.cm} | R{1.7cm} R{1.7cm} R{1.7cm} | R{1.7cm} R{1.7cm} R{1.7cm} | }
			\hline
			& \multicolumn{3}{c|}{\textbf{End-diastole}} & \multicolumn{3}{c|}{\textbf{End-systole}} \\
			\textbf{Scenario} & \multicolumn{1}{l}{\textbf{LV}} & \multicolumn{1}{l}{\textbf{RV}} & \multicolumn{1}{l|}{\textbf{LVM}} & \multicolumn{1}{l}{\textbf{LV}} & \multicolumn{1}{l}{\textbf{RV}} & \multicolumn{1}{l|}{\textbf{LVM}} \\ 
			\hline
			auto-only  &  \phantom{x}5.6$\pm$3.3 & \phantom{x}15.7$\pm$9.7 & \phantom{x}8.5$\pm$6.4 & \phantom{x}9.2$\pm$5.8 & \phantom{x}16.5$\pm$8.8 & \phantom{x}13.4$\pm$10.5 \\
			auto$+$simulation &  \phantom{x}4.5$\pm$2.1 & *\phantom{x}9.0$\pm$4.6 & *5.9$\pm$3.4 & *5.2$\pm$2.5 & *10.3$\pm$3.7 & *\phantom{x}6.6$\pm$2.9 \\
			auto$+$expert &  \phantom{x}4.9$\pm$2.8 & *\phantom{x}9.8$\pm$4.3 & \phantom{x}7.3$\pm$4.3 & \phantom{x}7.2$\pm$3.3 & *12.5$\pm$4.7 & *\phantom{x}8.3$\pm$3.5 \\
			
				\hline
		\end{tabular}
		\label{table_manual_seg_perf_hd}
	}   
	
	\subfloat[\textbf{Clinical metrics:} a) Left ventricle (LV) end-diastolic volume (EDV) b) LV ejection fraction (EF) c) Right ventricle (RV) EDV d) RV ejection fraction e) LV myocardial mass. Quantitative results compare clinical metrics based on reference segmentations with 1) automatic segmentations; 2) simulated manual correction and 3) manual expert correction of automatic segmentations using spatial uncertainty maps. $\rho$ denotes the Pearson correlation coefficient, \textit{bias} denotes the mean difference between the two measurements (mean $\pm$ standard deviation) and \textit{MAE} denotes the mean absolute error between the two measurements.]{
		\label{table_manual_cardiac_function_indices}
		\small
		\begin{tabular}{| C{2.cm} | C{0.5cm}  C{0.8cm} C{0.55cm} | C{0.5cm}  C{0.8cm}C{0.55cm} | C{0.5cm}  C{0.8cm} C{0.55cm} | C{0.5cm}  C{0.8cm} C{0.55cm} | C{0.5cm} C{0.8cm} C{0.55cm} |}
			\hline
			& \multicolumn{3}{c}{\textbf{LV$_{EDV}$}} &  \multicolumn{3}{c}{\textbf{LV$_{EF}$}} & \multicolumn{3}{c}{\textbf{RV$_{EDV}$}} &  \multicolumn{3}{c}{\textbf{RV$_{EF}$}} & \multicolumn{3}{c|}{\textbf{LVM$_{Mass}$}}  \\
			
			\textbf{Scenario} & \textbf{$\rho$} & \textbf{bias $\pm\sigma$} & \textbf{MAE} & \textbf{$\rho$} & \textbf{bias $\pm\sigma$} & \textbf{MAE} & \textbf{$\rho$} & \textbf{bias $\pm\sigma$} & \textbf{MAE} & \textbf{$\rho$} & \textbf{bias $\pm\sigma$} & \textbf{MAE} & \multicolumn{1}{c}{\textbf{$\rho$}} & \textbf{bias $\pm\sigma$} & \textbf{MAE} \\
			\hline
			
			auto-only  & 0.995 & -4.4 $\pm$7.0 & 5.7 & 0.927 & 5.0 $\pm$7.1 & 5.8 & 0.962 & -6.4 $\pm$16.2 & 11.9 & 0.878 & 5.8 $\pm$8.7 & 8.0 & 0.979 & -6.4 $\pm$10.6 & 9.5 \\
			
			auto$+$simulation & 0.998 & -3.9 $\pm$5.2 & 4.8 & 0.989 & 2.3 $\pm$2.9 & 2.9 & 0.984 & -3.7 $\pm$10.4 & 6.8 & 0.954 & 2.7 $\pm$5.5 & 4.5 & 0.983 & -5.5 $\pm$9.6 & 8.1 \\
			
			auto$+$expert & 0.996 & -4.3 $\pm$6.5 & 5.5 & 0.968 & 2.7 $\pm$4.8 & 4.3 & 0.976 & -3.2 $\pm$12.9 & 8.3 & 0.883 & 5.1 $\pm$8.6 & 7.7 & 0.980 & -6.2 $\pm$10.2 & 9.1 \\
				\hline
		\end{tabular}
	}
	
\end{table*}

\vspace{1ex}
\noindent \textbf{Detection of slices with segmentation failures}: To evaluate detection performance w.r.t. slices containing segmentation failures precision-recall curves for each combination of model architecture and loss function using e-maps (Figure~\ref{fig_prec_rec_slice_detection}, left) or b-maps (Figure~\ref{fig_prec_rec_slice_detection}, right) are shown. The results show that detection performance of slices containing segmentation failures is slightly better for all models when using e-maps. Furthermore, the detection network achieves highest performance using uncertainty maps obtained from the DN model and the lowest when exploiting e- or b-maps obtained from the DRN model. Table~\ref{table_evaluation_slice_detection} shows the average precision of detected slices with segmentation failures per patient, as well as the average percentage of slices that do contain segmentation failures (reference for detection task). The results illustrate that these measures are positively correlated i.e. that precision of detected slices in a patient volume is higher if the volume contains more slices that need correction. On average the DN model generates cardiac segmentations that contain more slices with at least one segmentation failure compared to U-net (ranks second) and DRN (ranks third). A higher number of detected slices containing segmentation failures implies an increased workload for manual correction.

\subsection{Calibration of uncertainty maps} \label{result_eval_quality_umaps}

\begin{figure*}[t]
	\centering
	\includegraphics[width=4.7in, height=2.8in]{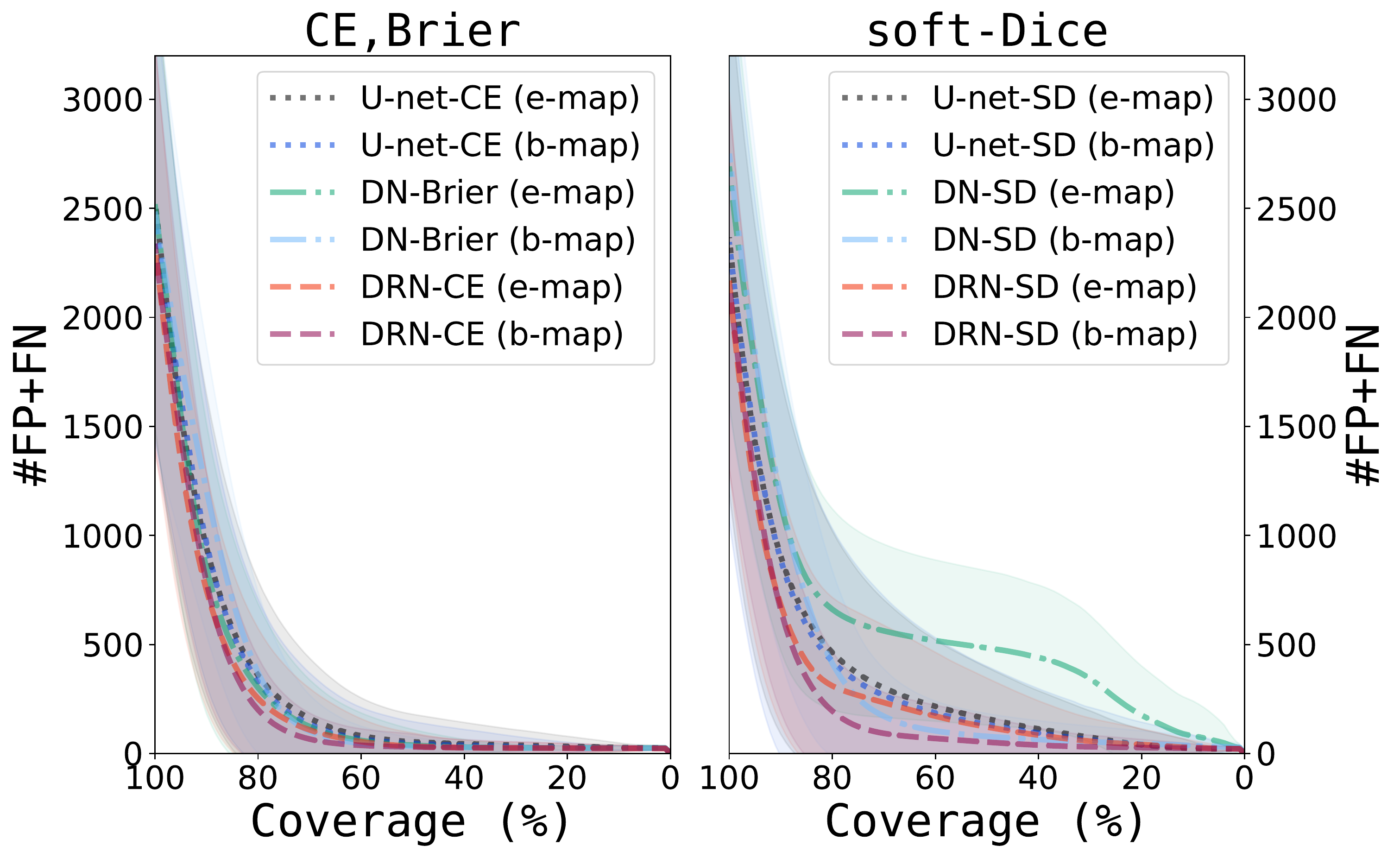}
	
	\caption{Comparison of risk-coverage curves for different combination of model architectures, loss functions and uncertainty maps. Results are separated for loss functions (left cross-entropy and Brier, right soft-Dice loss). \num{100}\% coverage means that none of the voxels is discarded based on its uncertainty whereas a coverage of \num{0}\% denotes the scenario in which all predictions are replaced by their reference labels. Note, all models were trained with two different loss functions (1) soft-Dice (SD) for all models (2) cross-entropy (CE) for DRN and U-net and Brier loss for DN.}
	\label{fig_risk_cov_comparison}
\end{figure*}

Figure~\ref{fig_risk_cov_comparison} shows risk-coverage curves for each combination of model architectures, uncertainty maps and loss functions (Figure~\ref{fig_risk_cov_comparison} left: CE or Brier loss, Figure~\ref{fig_risk_cov_comparison} right: soft-Dice). The results show an effect of the loss function on slope and convergence of the curves. Segmentation errors of models trained with the soft-Dice loss are less frequently covered by higher uncertainties than models trained with CE or Brier loss (steeper slope and lower minimum are better). This difference is more pronounced for e-maps. Models trained with the CE or Brier loss only slightly differ concerning convergence and their slopes are approximately identical. In contrast, the curves of the models trained with the soft-Dice differ regarding their slope and achieved minimum. Comparing e- and b-map of the DN-SD and U-net-SD models the results reveal that the curve for b-map has a steeper slope and achieves a lower minimum compared to the e-map. For the DRN-SD model these differences are less striking. In general for a specific combination of model and loss function the risk-coverage curves using b-maps achieve a lower minimum compared to e-maps.

\begin{table*}
	\caption{Effect of number of Monte Carlo (MC) samples on segmentation performance in terms of (a) Dice coefficient (DC) and (b) Hausdorff Distance (HD) (mean $\pm$ standard deviation). Higher DC and lower HD is better. Abbreviations: Cross-Entropy (CE), Dilated Residual Network (DRN) and Dilated Network (DN).} 
	\label{table_seg_perf_per_samples}
	\small
	\centering
	\subfloat[Dice coefficient]{
		\begin{tabular}{|c  C{1.5cm}  C{1.5cm} c|}
			\hline
			\textbf{\thead{Number of \\ MC samples}} & DRN-CE & U-net-CE & DN-soft-Dice\\
			\hline
			1 & 0.894$\pm$0.07 & 0.896$\pm$0.07 & 0.871$\pm$0.09 \\
			3 & 0.900$\pm$0.07 & 0.901$\pm$0.07 & 0.883$\pm$0.08 \\
			5 & 0.902$\pm$0.07 & 0.901$\pm$0.07 & 0.887$\pm$0.08 \\
			7 & 0.903$\pm$0.07 & 0.901$\pm$0.07 & 0.888$\pm$0.08 \\
			10 & 0.904$\pm$0.06 & 0.902$\pm$0.07 &0.890$\pm$0.08  \\
			20 & 0.904$\pm$0.07 & 0.902$\pm$0.07 & 0.890$\pm$0.08 \\
			30 & 0.904$\pm$0.07 & 0.902$\pm$0.07 & 0.891$\pm$0.08 \\
			60 & 0.904$\pm$0.07 & 0.902$\pm$0.07 & 0.891$\pm$0.08 \\
			\hline
		\end{tabular}
	}
	\subfloat[Hausdorff Distance]{
		\begin{tabular}{|c  C{1.5cm}  C{1.5cm} c|}
			\hline
			\textbf{\thead{Number of \\ MC samples}}& DRN-CE & U-net-CE & DN-soft-Dice \\
			\hline
			1 & 9.88$\pm$5.76 & 11.79$\pm$8.23 & 13.54$\pm$7.14 \\
			3 & 9.70$\pm$6.13 & 11.40$\pm$7.78 & 12.71$\pm$6.79 \\
			5 & 9.54$\pm$6.07 & 11.37$\pm$7.81 & 12.06$\pm$6.29 \\
			7 & 9.38$\pm$5.86 & 11.29$\pm$7.86 & 12.08$\pm$6.38 \\
			10 & 9.38$\pm$5.91 & 11.24$\pm$7.71 & 11.85$\pm$6.34 \\
			20 & 9.37$\pm$5.83 & 11.27$\pm$7.79 & 11.90$\pm$6.52 \\
			30 & 9.39$\pm$5.91 &  11.32$\pm$7.93 & 11.90$\pm$6.48 \\
			60 & 9.39$\pm$5.93 & 11.22$\pm$7.83 & 11.89$\pm$6.56 \\
			\hline
		\end{tabular}	
	}
\end{table*}

\subsection{Correction of automatically identified segmentation failures} \label{results_combined_approach}

\textbf{Simulated correction:} The results listed in Table~\ref{table_overall_segmentation_performance} and \ref{table_cardiac_function_indices} show that the proposed method consisting of segmentation followed by simulated manual correction of detected segmentation failures delivers accurate segmentation for all tissues over ED and ES points. Correction of detected segmentation failures improved the performance in terms of DC, HD and clinical metrics for all combinations of model architectures, loss functions and uncertainty measures. Focusing on the DC after correction of detected segmentation failures the results reveal that performance differences between evaluated models decreased compared to the segmentation-only task. This effect is less pronounced for HD where the DRN network clearly achieved superior results in the segmentation-only and combined approach. The DN performs the least of all models but achieves the highest absolute DC performance improvements in the combined approach for RV at ES. Overall, the results in Table~\ref{table_overall_segmentation_performance} disclose that improvements attained by the combined approach are almost all statistically significant ($p \leq 0.05$) at ES and frequently at ED (\num{96}\% resp. \num{83}\% of the cases). Moreover, improvements are in \num{99}\% of the cases statistically significant for HD compared to \num{81}\% of the cases for DC.

Results in terms of clinical metrics shown in Table~\ref{table_cardiac_function_indices} are inline with these findings. We observe that segmentation followed by simulated manual correction of detected segmentation failures resulted in considerably higher accuracy for LV and RV ejection fraction. Achieved improvements for clinical metrics are only statistically significant ($p \leq 0.05$) in one case for RV ejection fraction.	

In general, the effect of correction of detected segmentation failures is more pronounced in cases where the segmentation-only approach achieved relatively low accuracy (e.g. DN-SD for RV at ES). Furthermore, performance gains are largest for RV and LV at ES and for ejection fraction of both ventricles.

\begin{figure*}[!t]
	\captionsetup[subfigure]{justification=centering}
	\centering
	\subfloat[]{\includegraphics[width=6in]{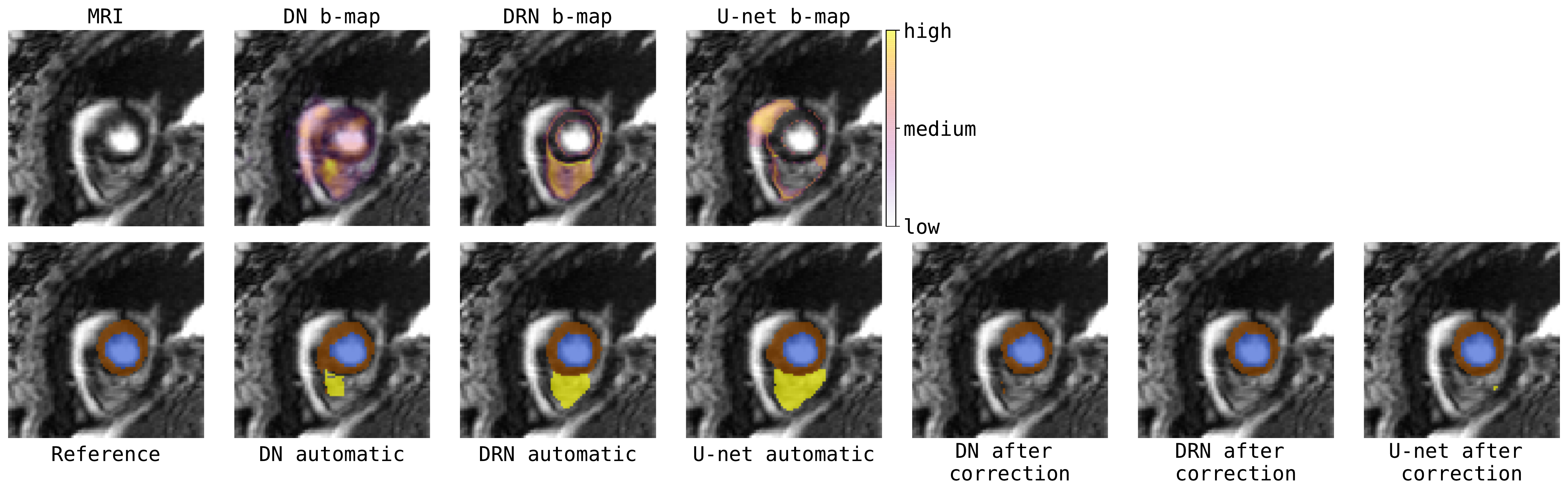}%
		\label{fig_qual_result_sim_corr_example1}}
	
	\subfloat[]{\includegraphics[width=6in]{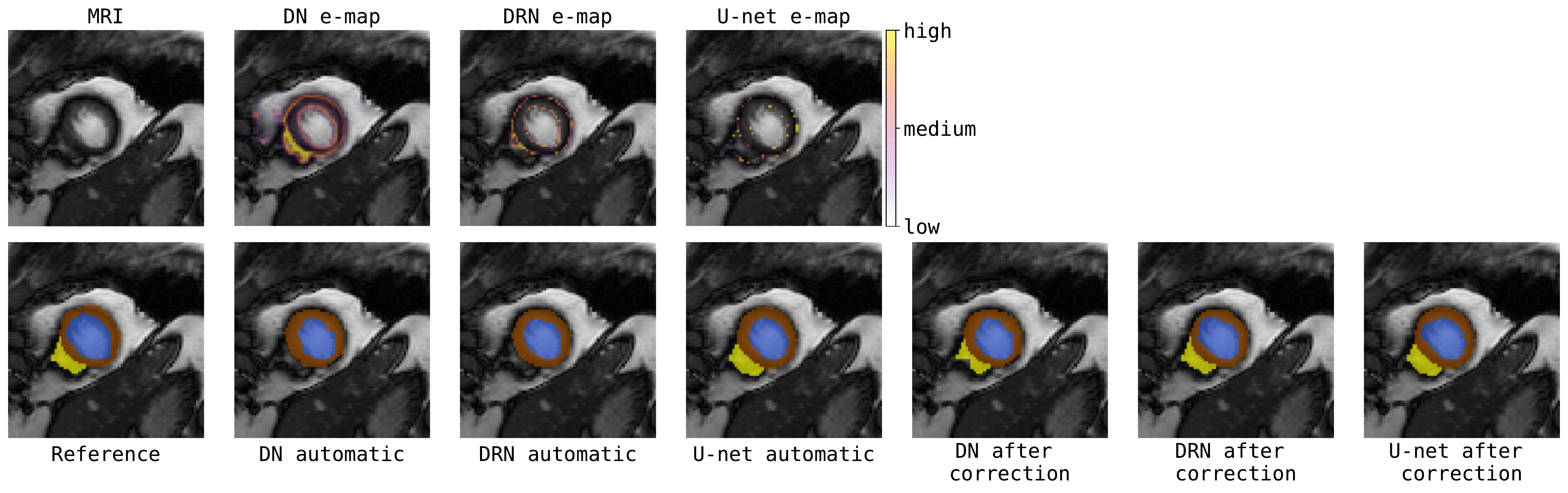}%
		\label{fig_qual_result_sim_corr_example2}}
	
	\subfloat[]{\includegraphics[width=6in]{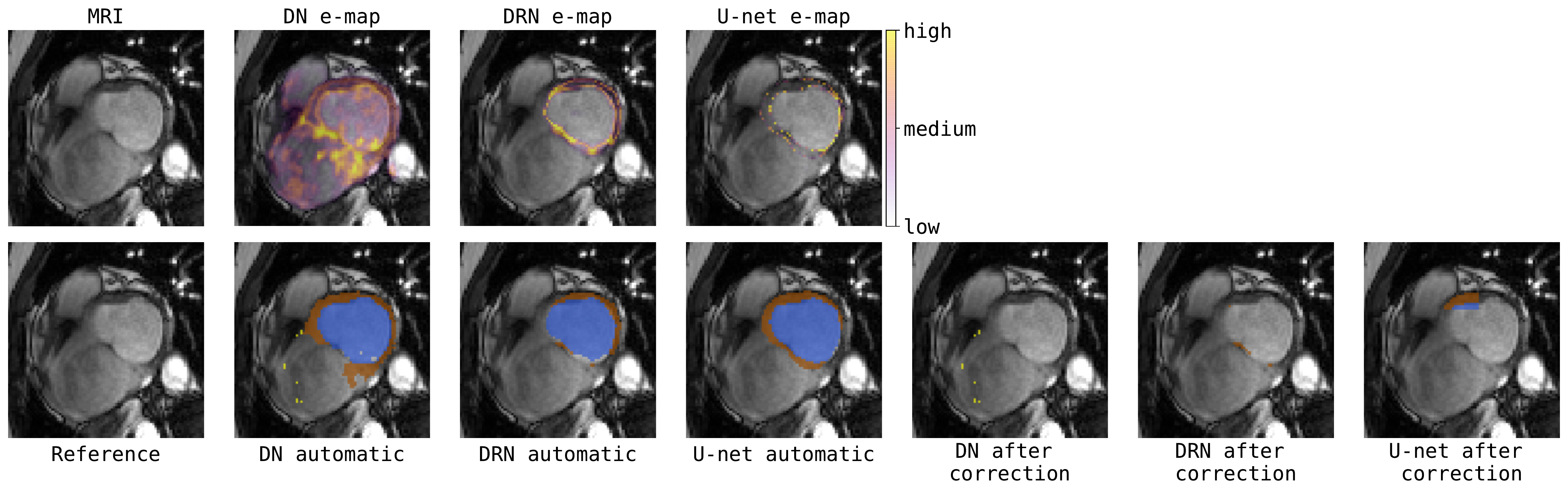}%
		\label{fig_qual_result_sim_corr_example3}}
	\caption{Three patients showing results of combined segmentation and detection approach consisting of segmentation followed by simulated manual correction of detected segmentation failures. First column shows MRI (top) and reference segmentation (bottom). Results for automatic segmentation and simulated manual correction respectively achieved by: Dilated Network (DN-Brier, \num{2}$^{nd}$ and \num{5}$^{th}$ columns); Dilated Residual Network (DRN-soft-Dice, \num{3}$^{rd}$ and \num{6}$^{th}$ columns); and U-net (soft-Dice, \num{4}$^{th}$ and \num{7}$^{th}$ columns).}
	\label{fig_seg_detection_qualitative_results}
\end{figure*}

\begin{figure*}[!t]
	\captionsetup[subfigure]{justification=centering}
	\centering
	\subfloat[]{\includegraphics[width=3.3in]{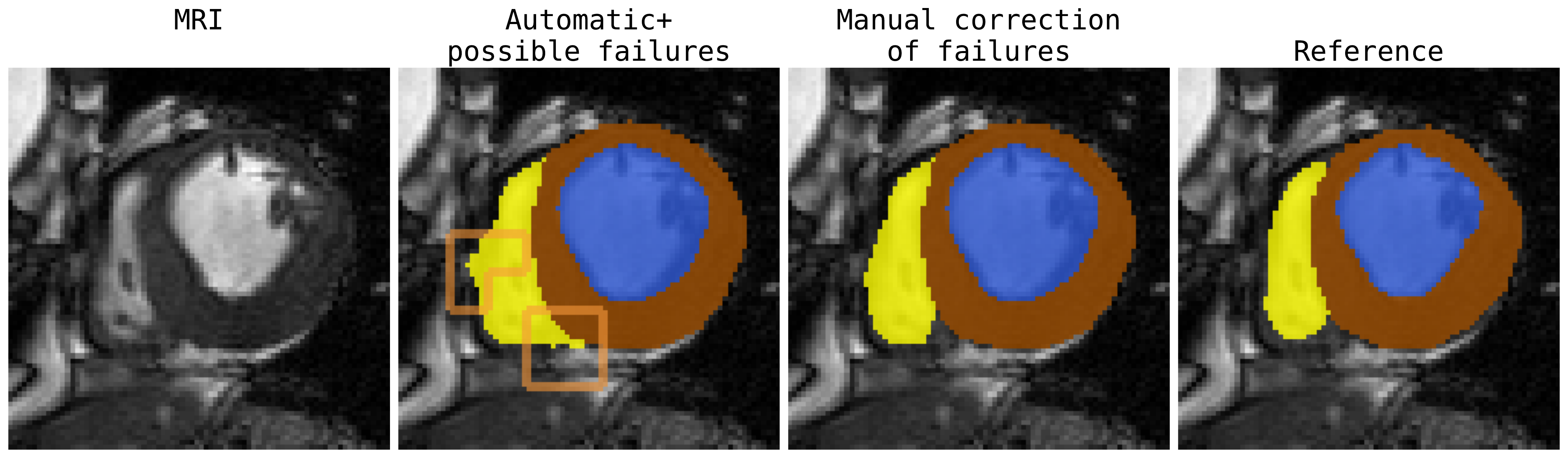}%
		\label{fig_qual_result_man_corr_example1} \hspace{3ex}}
	\subfloat[]{\includegraphics[width=3.3in]{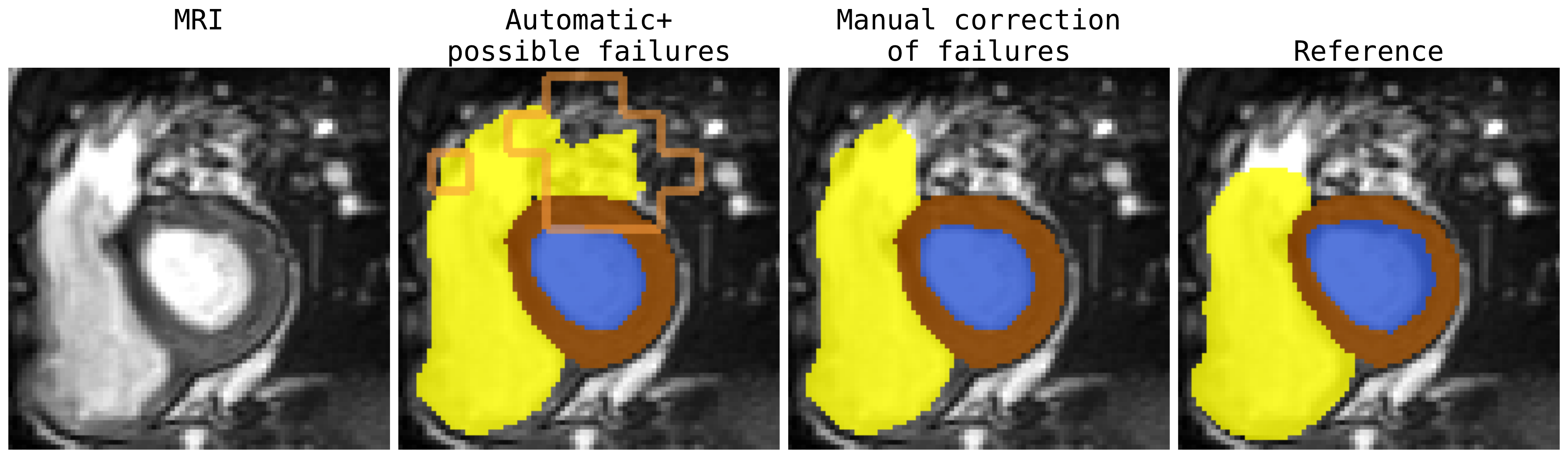}%
		\label{fig_qual_result_man_corr_example2}}
	
	\subfloat[]{\includegraphics[width=3.3in]{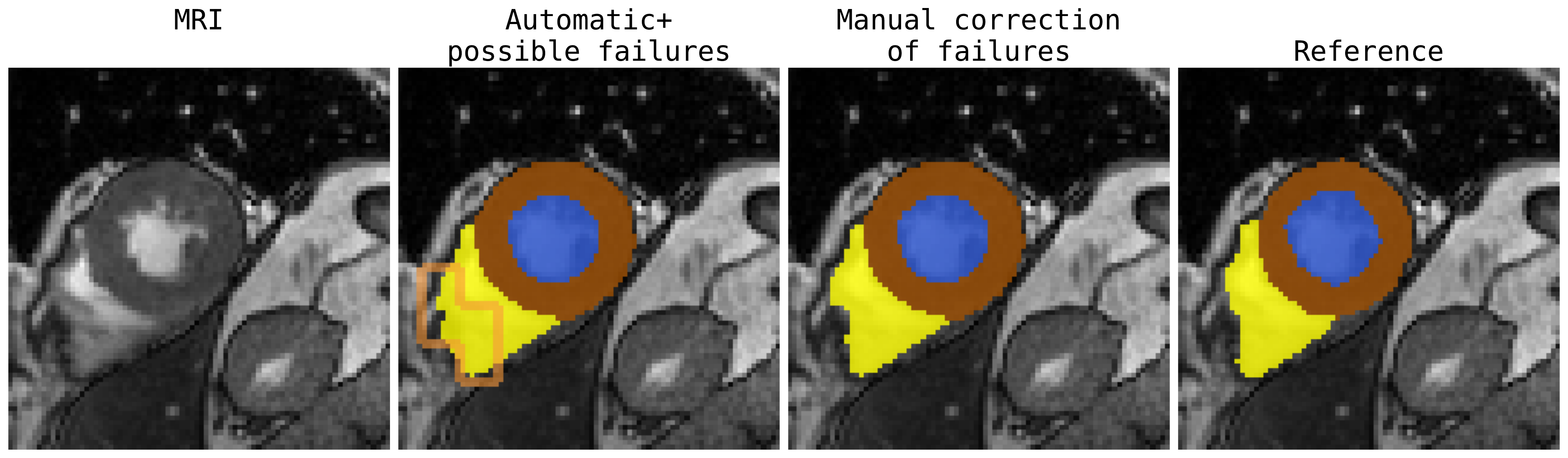}%
		\label{fig_qual_result_man_corr_example3} \hspace{3ex}}
	\subfloat[]{\includegraphics[width=3.3in]{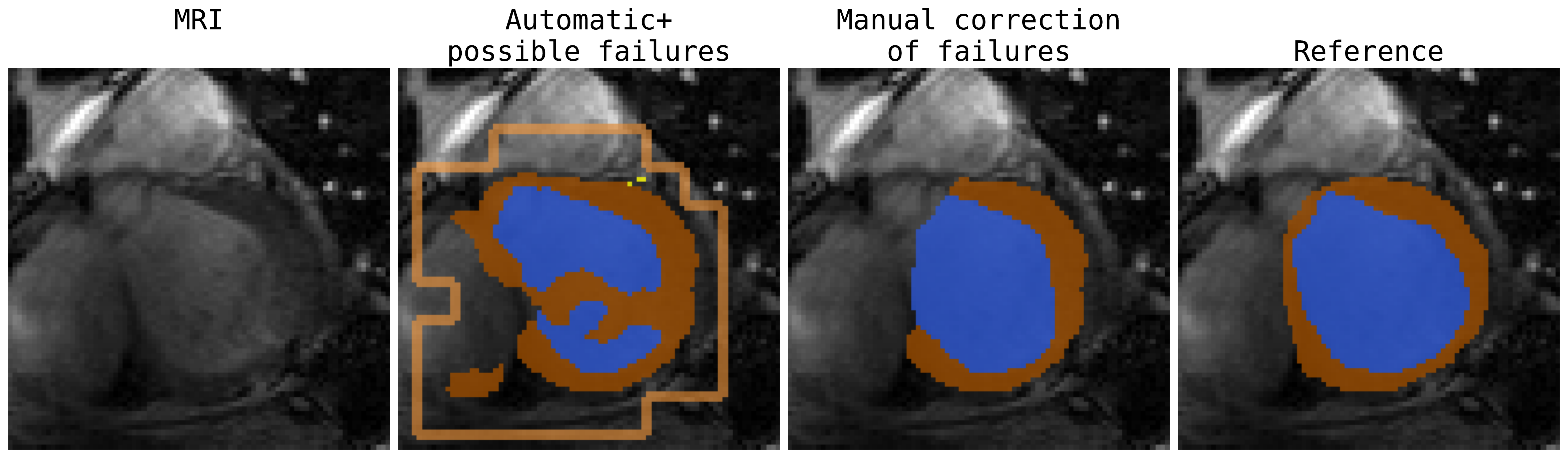}%
		\label{fig_qual_result_man_corr_example4}}  
	
	\caption{Four patients showing results of combined segmentation and detection approach consisting of segmentation followed by manual expert correction of detected segmentation failures. Expert was only allowed to adjust the automatic segmentations in regions where the detection network predicted segmentation failures (orange contour shown in 2$^{nd}$ column). Automatic segmentations were generated by a U-net trained with the cross-entropy loss. Segmentation failure detection was performed using entropy maps.}
	\label{fig_qualitative_results_man_corr}
\end{figure*}

The best overall performance is achieved by the DRN model trained with cross-entropy loss while exploiting entropy maps in the detection task. Moreover, the proposed two step approach attained slightly better results using Bayesian maps compared to entropy maps. 

\vspace{1ex}
\textbf{Manual correction}: Table~\ref{table_manual_corr_performance} lists results for the combined automatic segmentation and detection approach followed by \textit{manual} correction of detected segmentation failures by an expert. The results show that this correction led to improved segmentation performance in terms of DC, HD and clinical metrics. Improvements in terms of HD are in \num{50} percent of the cases statistically significant ($p \leq 0.05$) and most pronounced for RV and LV at end-systole. 

Qualitative examples of the proposed approach are visualized in Figures~\ref{fig_seg_detection_qualitative_results} and \ref{fig_qualitative_results_man_corr} for simulated correction and manual correction of the automatically detected segmentation failures, respectively. For the illustrated cases (simulated) manual correction of detected segmentation failures leads to increased segmentation performance.
On average manual correction of automatic segmentations took less than \num{2} minutes for ED and ES volumes of one patient compared to \num{20} minutes that is typically needed by an expert for the same task.

\section{Ablation Study}

To demonstrate the effect of different hyper-parameters in the method, a number of experiments were performed. In the following text these are detailed.

\subsection{Impact of number of Monte Carlo samples on segmentation performance}

To investigate the impact of the number of Monte Carlo (MC) samples on the segmentation performance validation experiments were performed for all three segmentation architectures (Dilated Network, Dilated Residual Network and U-net) using $T$ $\in \{1, 3, 5, 7, 10, 20, 30, 60\}$ samples. Results of these experiments are listed in Table~\ref{table_seg_perf_per_samples}. We observe that segmentation performance started to converge using \num{7} samples only. Performance improvements using an increased number of MC samples were largest for the Dilated Network. Overall, using more than \num{10} samples did not increase segmentation performance. Hence, in the presented work $T$ was set to \num{10}.	

\subsection{Effect of patch-size on detection performance}

\begin{figure*}[t]
	\center
	\subfloat[]{\includegraphics[width=3.4in, height=1.7in]{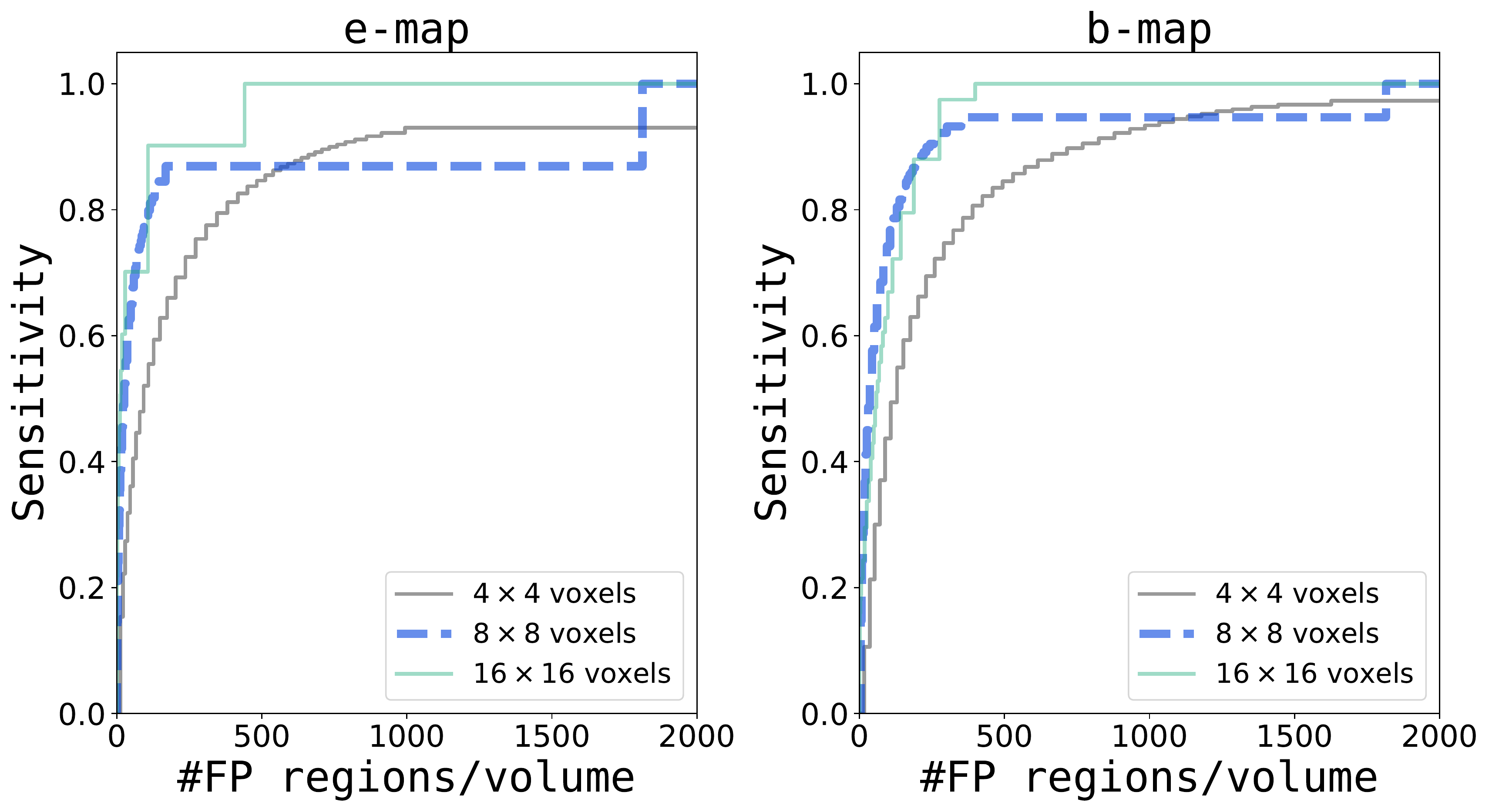}%
		\label{fig_fn_grid_froc_voxel_detection}}
	\subfloat[]{\includegraphics[width=3.4in, height=1.7in]{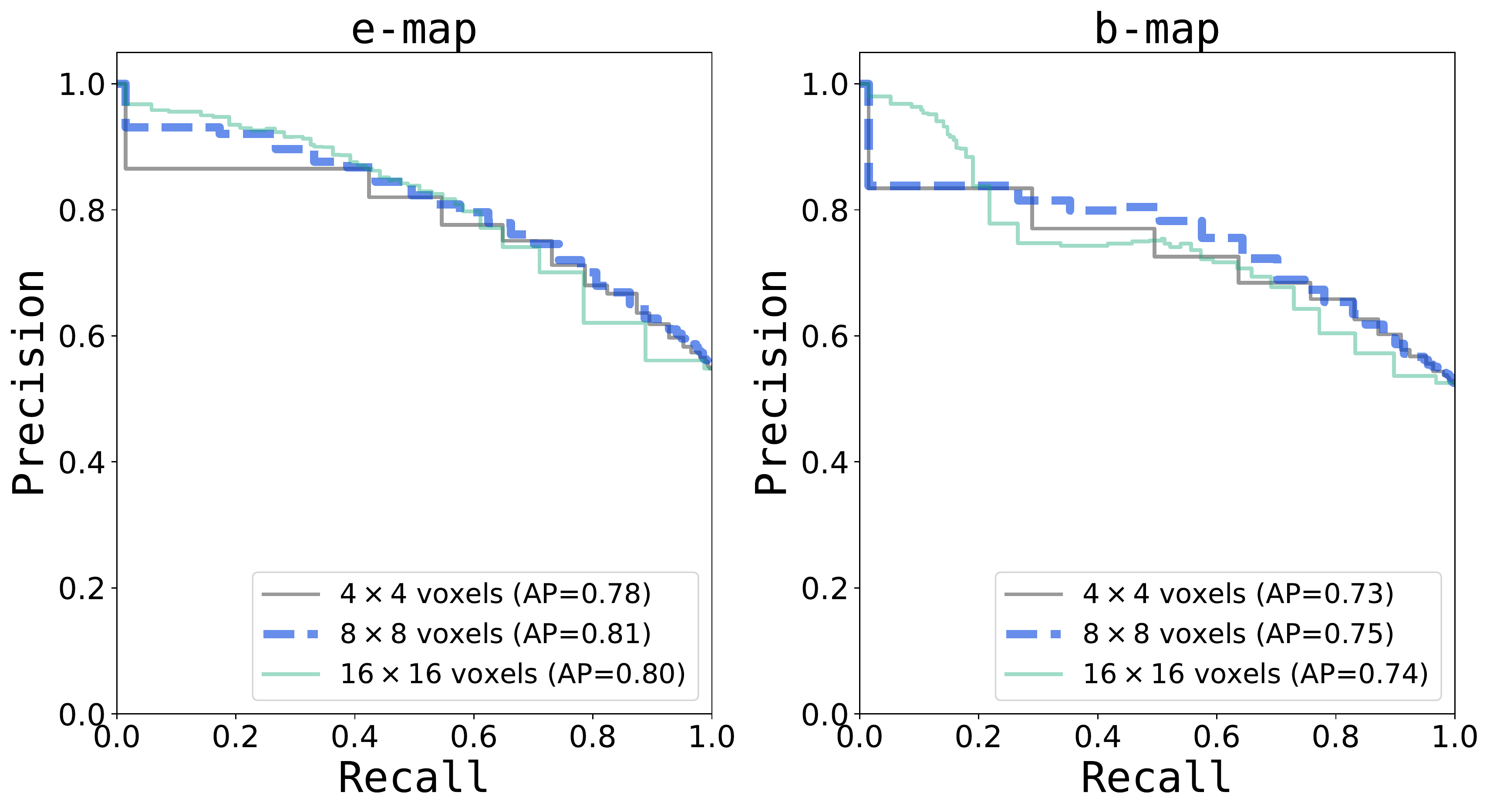}%
		\label{fig_fn_grid_prec_rec_slice_detection}} 
	
	\caption{Detection performance for three different patch-sizes specified in voxels. (a) Sensitivity for detection of segmentation failures on voxel level (y-axis) versus number of false positive image regions (x-axis). (b) Precision-recall curve for detection of slices containing segmentation failures (where AP denotes average precision). Results are split between entropy and Bayesian uncertainty maps. In the experiments patch-size was set to \num{8}$\times$\num{8} voxels.}
	\label{fig_grid_compare}
\end{figure*}

The combined segmentation and detection approach detects segmentation failures on region level. To investigate the effect of patch-size on detection performance three different patch-sizes were evaluated: \num{4}$\times$\num{4}, \num{8}$\times$\num{8}, and \num{16}$\times$\num{16} voxels. The results are shown in Figure~\ref{fig_grid_compare}. We can observe in Figure~\ref{fig_fn_grid_froc_voxel_detection} that larger patch-sizes result in a lower number of false positive regions. The result is potentially caused by the decreasing number of regions in an image when using larger patch-sizes compared to smaller patch-sizes. Furthermore, Figure~\ref{fig_fn_grid_prec_rec_slice_detection} reveals that slice detection performance is only slightly influenced by patch-size. To ease manual inspection and correction by an expert, it is desirable to keep region-size i.e. patch-size small. Therefore, in the experiments a patch-size of \num{8}$\times$\num{8} voxels was used.

\subsection{Impact of tolerance threshold on number of segmentation failures}

To investigate the impact of the tolerance threshold separating segmentation failures and tolerable segmentation errors, we calculated the ratio of the number of segmentation failures and all errors i.e. the sum of tolerable errors and segmentation failures. Figure~\ref{fig_threshold_compare} shows the results. We observe that at least half of the segmentation failures are located within a tolerance threshold i.e. distance of two to three voxels of the target structure boundary as defined by the reference annotation. Furthermore, the mean percentage of failures per volume is considerably lower for the Dilated Residual Network (DRN) and highest for the Dilated Network. This result is inline with our earlier finding (see Table~\ref{table_evaluation_slice_detection}) that average percentage of slices that do contain segmentation failures is lowest for the DRN model.

\section{Discussion}

We have described a method that combines automatic segmentation and assessment of uncertainty in cardiac MRI with detection of image regions containing segmentation failures. The results show that combining automatic segmentation with manual correction of detected segmentation failures results in higher segmentation performance. 
In contrast to previous methods that detected segmentation failures per patient or per structure, we showed that it is feasible to detect segmentation failures per image region. In most of the experimental settings, simulated manual correction of detected segmentation failures for LV, RV and LVM at ED and ES led to statistically significant improvements. These results represent the upper bound on the maximum achievable performance	for the manual expert correction task. Furthermore, results show that manual expert correction of detected segmentation failures led to consistently improved segmentations. However, these results are not on par with the simulated expert correction scenario. This is not surprising because inter-observer variability is high for the presented task and annotation protocols may differ between clinical environments. Moreover, qualitative results of the manual expert correction reveal that manual correction of the detected segmentation failures can prevent anatomically implausible segmentations (see Figure~\ref{fig_qualitative_results_man_corr}).
Therefore, the presented approach can potentially simplify and accelerate correction process and has the capacity to increase the trustworthiness of existing automatic segmentation methods in daily clinical practice.

The proposed combined segmentation and detection approach was evaluated using three state-of-the-art deep learning segmentation architectures. The results suggest that our approach is generic and applicable to different model architectures. Nevertheless, we observe noticeable differences between the different combination of model architectures, loss functions and uncertainty measures. In the segmentation-only task the DRN clearly outperforms the other two models in the evaluation of the boundary of the segmented structure. Moreover, qualitative analysis of the automatic segmentation masks suggests that DRN generates less often anatomically implausible and fragmented segmentations than the other models. We assume that clinical experts would prefer such segmentations although they are not always perfect. Furthermore, even though DRN and U-net achieve similar performance in regard to DC we assume that less fragmented segmentation masks would increase trustworthiness of the methods. 

\begin{figure*}[t]
	\center
	\subfloat[]{
		\includegraphics[width=3.in]{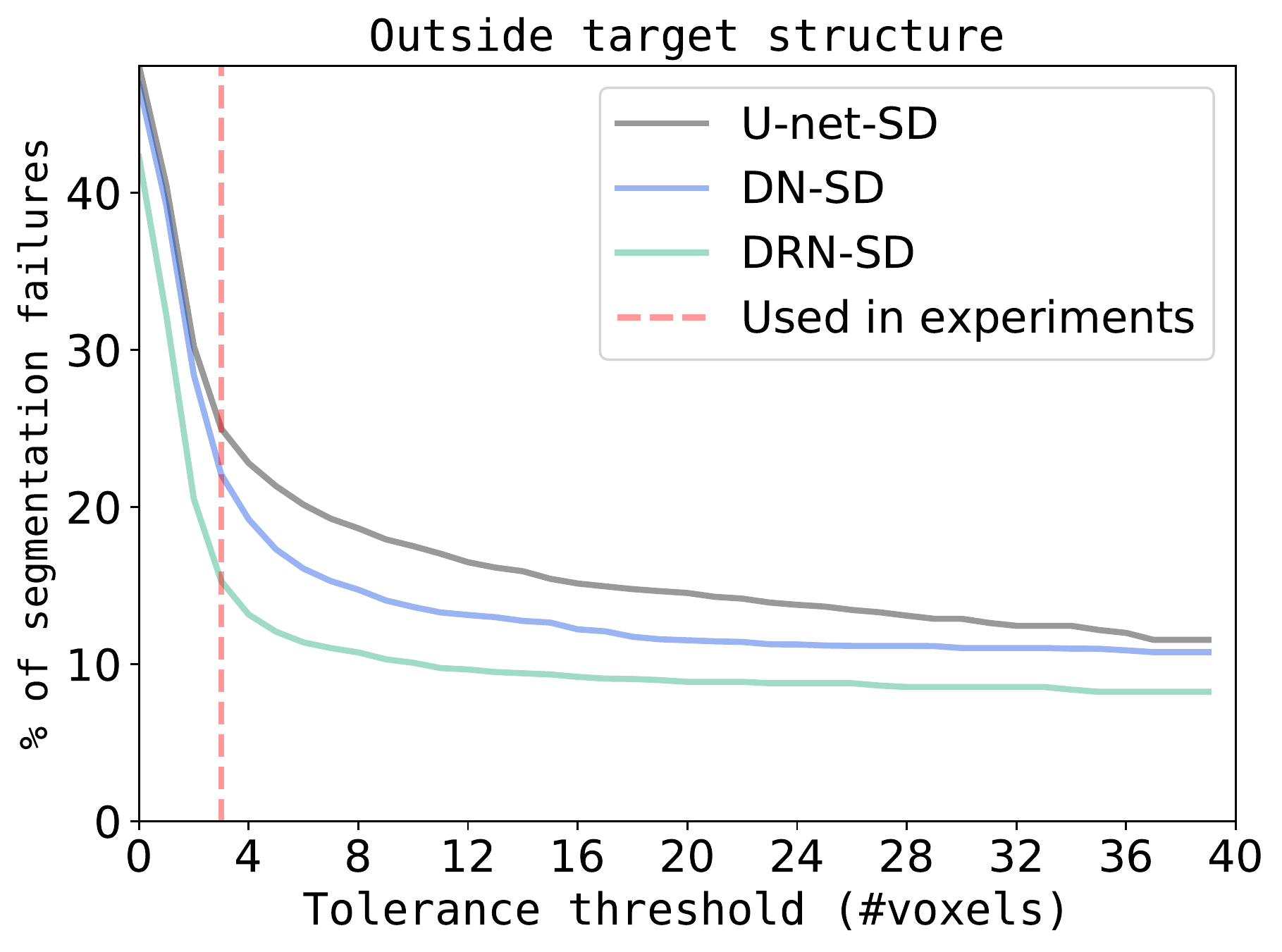}%
		\label{fig_threshold_struc_out_compare}
	}
	\subfloat[]{
		\includegraphics[width=3.in]{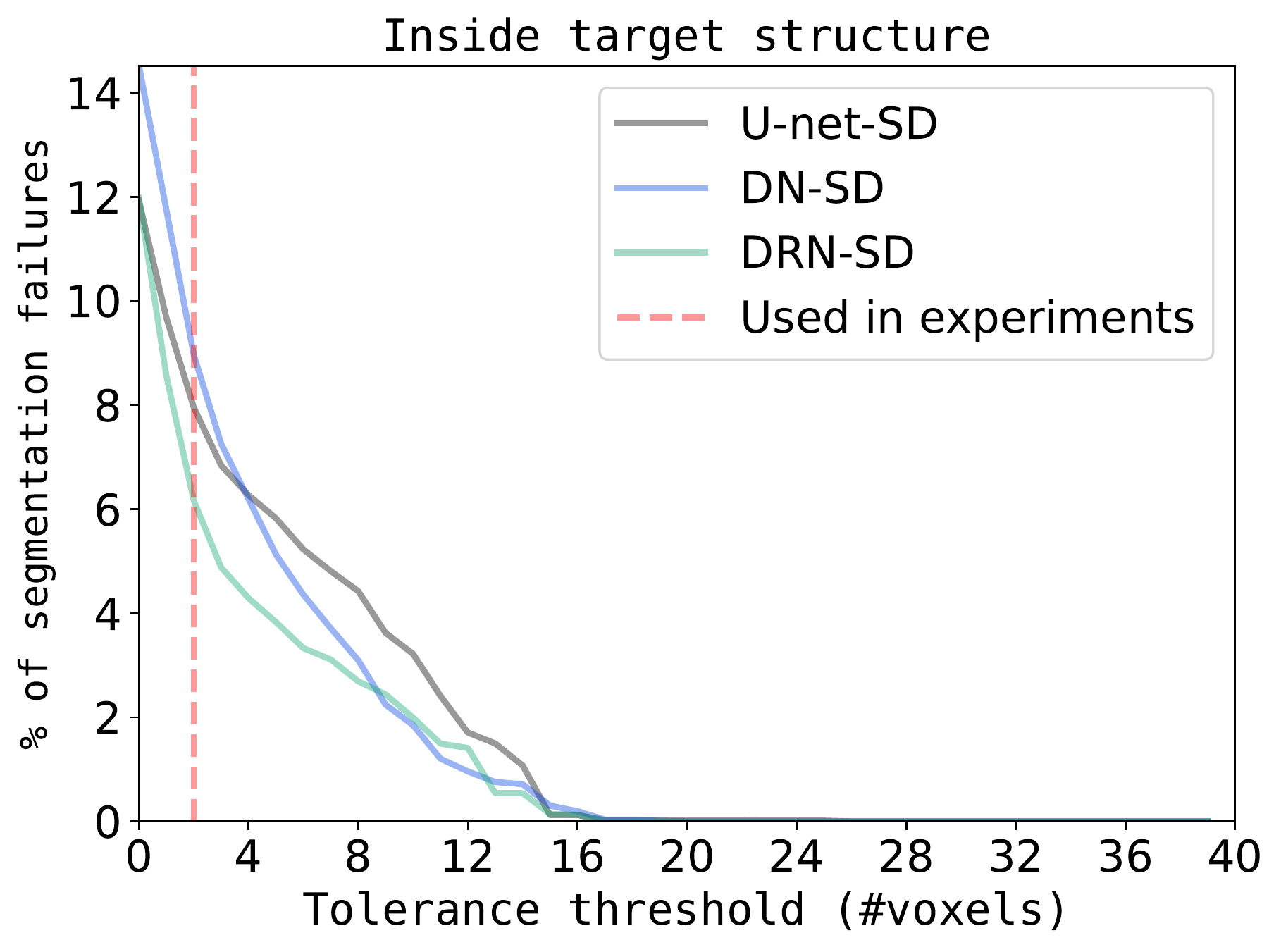}%
		\label{fig_threshold_struc_in_compare}
	}
	\caption{Mean percentage of the segmentation failures per volume (y-axis) in the set of all segmentation errors (tolerable errors$+$segmentation failures) depending on the tolerance threshold (x-axis). Red, dashed vertical line indicates threshold value that was used throughout the experiments. Results are split between segmentation errors located (a) outside and (b) inside the target structure. Each figure contains a curve for U-net, Dilated Network (DN) and Dilated Residual Network (DRN) trained with the soft-Dice (SD) loss. Segmentation errors located in slices above the base or below the apex are always included in the set of segmentation failures and therefore, they are independent of the applied tolerance threshold.}
	\label{fig_threshold_compare}
\end{figure*}

In agreement with our preliminary work we found that uncertainty maps obtained from a segmentation model trained with soft-Dice loss have a lower degree of uncertainty calibration compared to when trained with one of the other two loss functions (cross-entropy and Brier)\cite{sander2019towards}. Nevertheless, the results of the combined segmentation and detection approach showed that a lower degree of uncertainty calibration only slightly deteriorated the detection performance of segmentation failures for the larger segmentation models (DRN and U-net) when exploiting uncertainty information from e-maps. Hendrycks and Gimpel \cite{hendrycks2016baseline} showed that softmax probabilities generated by deep learning networks have poor direct correspondence to confidence. However, in agreement with Geifman et al. \cite{geifman2017selective} we presume that probabilities and hence corresponding entropies obtained from softmax function are ranked consistently i.e. entropy can potentially be used as a relative uncertainty measure in deep learning. In addition, we detect segmentation failures per image region and therefore, our approach does not require perfectly calibrated uncertainty maps. Furthermore, results of the combined segmentation and detection approach revealed that detection performance of segmentation failures using b-maps is almost independent of the loss function used to train the segmentation model. In line with Jungo et al. \cite{jungo2019assessing} we assume that by enabling MC-dropout in testing and computing the mean softmax probabilities per class leads to better calibrated probabilities and b-maps. This assumption is in agreement with Srivastava et al.~\cite{srivastava2014dropout} where a CNN with dropout used at testing is interpreted as an ensemble of models.

Quantitative evaluation in terms of Dice coefficient and Hausdorff distance reveals that proposed combined segmentation and detection approach leads to significant performance increase. However, the results also demonstrate that the correction of the detected failures allowed by the combined approach does not lead to statistically significant improvement in clinical metrics. This is not surprising because state-of-the-art automatic segmentation methods are not expected to lead to large volumetric errors \cite{bernard2018deep} and standard clinical measures are not sensitive to small segmentation errors. Nevertheless, errors of the current state-of-the-art automatic segmentation methods may lead to anatomically implausible segmentations \cite{bernard2018deep} that may cause distrust in clinical application. Besides increase in trustworthiness of current state-of-the-art segmentation methods for cardiac MRIs, improved segmentations are a prerequisite for advanced functional analysis of the heart e.g. motion analysis\cite{bello2019deep} and very detailed morphology analysis such as myocardial trabeculae in adults\cite{meyer2020genetic}.

For the ACDC dataset used in this manuscript, Bernard et al.\cite{bernard2018deep} reported inter-observer variability ranging from \num{4} to \SI{14.1}{\milli\meter} (equivalent to on average \num{2.6} to \num{9} voxels). To define the set of segmentation failures, we employed a strict tolerance threshold on distance metric to distinguish between tolerated segmentation errors and segmentation failures (see Ablation study). Stricter tolerance threshold was used because the thresholding is performed in \num{2}D, while evaluation of segmentation is done in \num{3}D. Large slice thickness in cardiac MR  could lead to a discrepancy between the two. As a consequence of this strict threshold results listed in Table~\ref{table_evaluation_slice_detection} show that almost all patient volumes contain at least one slice with a segmentation failure. This might render the approach less feasible in clinical practice. Increasing the threshold decreases the number of segmentation failures and slices containing segmentation failures (see Figure~\ref{fig_threshold_compare}) but also lowers the upper bound on the maximum achievable performance. Therefore, to show the potential of our proposed approach we chose to apply a strict tolerance threshold. Nevertheless, we realize that although manual correction of detected segmentation failures leads to increased segmentation accuracy the performance of precision-recall is limited (see Figure~\ref{fig_dt_perf_all_models}) and hence, should be a focus of future work. 

The presented patch-based detection approach combined with (simulated) manual correction can in principle lead to stitching artefacts in the resulting segmentation masks. A voxel-based detection approach could potentially solve this. However, voxel-based detection methods are more challenging to train due to the very small number of voxels in an image belonging to the set of segmentation failures.

Evaluation of the proposed approach for \num{12} possible combinations of segmentation models (three), loss functions (two) and uncertainty maps (two) resulted in an extensive number of experiments. Nevertheless, future work could extend evaluation to other segmentation models, loss functions or combination of losses. Furthermore, our approach could be evaluated using additional uncertainty estimation techniques e.g. by means of ensembling of networks \cite{lakshminarayanan2017simple} or variational dropout \cite{kingma2015variational}. In addition, previous work by Kendall and Gal~\cite{kendall2017uncertainties}, Tanno et al. \cite{tanno2019uncertainty} has shown that the quality of uncertainty estimates can be improved if model (epistemic) and data (aleatoric) uncertainty are assessed simultaneously with separate measures. The current study focused on the assessment of model uncertainty by means of MC-dropout and entropy which is a combination of epistemic and aleatoric uncertainty. Hence, future work could investigate whether additional estimation of aleatoric uncertainty improves the detection of segmentation failures. 

Furthermore, to develop an end-to-end approach future work could incorporate the detection of segmentation failures into the segmentation network. Besides, adding the automatic segmentations to the input of the detection network could increase the detection performance. 

Finally, the proposed approach is not specific to cardiac MRI segmentation. Although data and task specific training would be needed the approach could potentially be applied to other image modalities and segmentation tasks.

\section{Conclusion}

A method combining automatic segmentation and assessment of segmentation uncertainty in cardiac MR with detection of image regions containing local segmentation failures has been presented. The combined approach, together with simulated and manual correction of detected segmentation failures, increases performance compared to segmentation-only. The proposed method has the potential to increase trustworthiness of current state-of-the-art segmentation methods for cardiac MRIs. 

\section*{Data and code availability}
All models were implemented using the PyTorch\cite{paszke2017automatic} framework and trained on one Nvidia GTX Titan X GPU with \num{12} GB memory. The code to replicate the study is publicly available at  \href{https://github.com/toologicbv/cardiacSegUncertainty}{https://github.com/toologicbv/cardiacSegUncertainty}.


\section*{Acknowledgements}

This study was performed within the DLMedIA program (P15-26) funded by Dutch Technology Foundation with participation of PIE Medical Imaging.

\section*{Author contributions statement}

J.S., B.D.V. and I.I. designed the concept of the study. J.S. conducted the experiments. J.S., B.D.V. and I.I. wrote the manuscript. All authors reviewed the manuscript. 

\section*{Additional information}

\textbf{Competing interests}: The authors declare that they have no competing interests. 

\ifCLASSOPTIONcaptionsoff
\newpage
\fi



%


\bibliographystyle{IEEEtran}
\bibliography{cardiacMRI_seg_uncertainty}





\end{document}